\documentclass[acmsmall]{acmart}

\AtBeginDocument{%
  \providecommand\BibTeX{{%
    \normalfont B\kern-0.5em{\scshape i\kern-0.25em b}\kern-0.8em\TeX}}}

\copyrightyear{2022}
\acmYear{2022}
\setcopyright{rightsretained}
\acmConference[SIGMETRICS '22]{SIGMETRICS '22}{June 6--10, 2022}{Mumbai, India}
\acmBooktitle{SIGMETRICS '22, June 6--10, 2022, Mumbai, India}\acmDOI{10.1145/3489048.3526954}
\acmISBN{978-1-4503-9141-2/22/06}

\usepackage[normalem]{ulem}\usepackage{datetime}

\usepackage{balance}  % to better equalize the last page
\usepackage{graphics} % for EPS, load graphicx instead
\usepackage{url}
\usepackage{hyperref}
\usepackage{txfonts}
\usepackage{pslatex}    
\usepackage{pifont}
\usepackage{graphicx}
\usepackage{caption}
\usepackage{multirow}
\usepackage{makecell}
\usepackage[justification=centering]{caption}
\usepackage{xspace}
\usepackage{comment}
\usepackage{listings}
\usepackage{tikz}
\usepackage{calc}
\usepackage{fancyvrb}
\usepackage{xcolor}
\usepackage{flushend}
\usepackage{wrapfig}

\usepackage{algorithm2e}
\usepackage{algorithmic}

\newcommand{\xmark}{\ding{55}}%

\newcommand{\CP}{{CachePerf}}

\newcommand*{\rom}[1]{\romannumeral #1}

\usepackage{ragged2e}

\begin{document}

%\title{CachePerf: Holistically Identify Cache-Related Performance Issues}
%\title{CachePerf: Ratio-Based Cache Miss Classification}
\title[CachePerf]{CachePerf: A Unified Cache Miss Classifier via Hybrid Hardware Sampling}\thanks{We have filed  a U.S. patent with the serial number 63/281,942. Please contact with tongping@umass.edu for the licence. }
% CachePerf: Accurate and Lightweight Cache Misses Identification
%Accurately and Comprehensively Pinpointing CachePerf: Identifying All Cache Performance Issues

\author{Jin Zhou}
\email{jinzhou@umass.edu}
\affiliation{%
  \institution{University of Massachusetts Amherst}
%   \city{Amherst}
%   \state{Massachusetts}
  \country{USA}
%   \postcode{0100}
}

\author{Steven (Jiaxun) Tang}
\email{jtang@umass.edu}
\affiliation{%
  \institution{University of Massachusetts Amherst}
%   \city{Amherst}
%   \state{Massachusetts}
  \country{USA}
%   \postcode{0100}
}

\author{Hanmei Yang}
\email{hanmeiyang@umass.edu}
\affiliation{%
  \institution{University of Massachusetts Amherst}
%   \city{Amherst}
%   \state{Massachusetts}
  \country{USA}
%   \postcode{0100}
}

\author{Tongping Liu}
\email{tongping@umass.edu}
\affiliation{%
  \institution{University of Massachusetts Amherst}
%   \city{Amherst}
%   \state{Massachusetts}
  \country{USA}
%   \postcode{0100}
}

\date{}
\begin{abstract}
The cache plays a key role in determining the performance of applications, no matter for sequential or concurrent
programs on homogeneous and heterogeneous architecture. Fixing cache misses requires to understand the origin
and the type of cache misses. However, this remains to be an unresolved issue even after decades of research.
This paper proposes a unified profiling tool--\CP{}--that could correctly identify different types of cache
misses, differentiate allocator-induced issues from those of applications, and
exclude minor issues without much performance impact. The core idea behind CachePerf is a hybrid sampling
scheme: it employs the PMU-based coarse-grained sampling to select very few susceptible instructions (with frequent cache misses) and then employs the breakpoint-based fine-grained sampling to collect the
memory access pattern of these instructions. Based on our evaluation, CachePerf only imposes 14\% performance
overhead and 19\% memory overhead (for applications with large footprints), while identifying the types of
cache misses correctly. CachePerf detected 9 previous-unknown bugs.
Fixing the reported bugs achieves from 3\% to 3788\% performance speedup. CachePerf will be an indispensable
complementary to existing profilers due to its effectiveness and low overhead.

\end{abstract}


\maketitle

\section{Introduction}
\label{sec:intro}

Cache accesses are typically orders of magnitude faster (e.g., $200\times$~\cite{misspenalty}) than memory accesses. Therefore, it is critical to reduce cache misses in order to boost the performance of applications, no matter for single-threaded or multi-threaded applications running on homogeneous or heterogeneous hardware architectures. However, it is challenging to identify cache misses statically, as they are related to access pattern~\cite{Sheriff}, hardware feature (e.g., cache capacity, cache line size), or even starting addresses of objects~\cite{Predator}.   

Many tools aiming to identify cache misses have been developed in the past. Simulation-based approaches, such as different Pin tools~\cite{jaleel2008cmp, MultiCacheSim}, or cachegrind (one tool inside Valgrind~\cite{Valgrind})~\cite{seward2004cachegrind},
typically impose prohibitive performance overhead (e.g., 100 times) that makes them even unsuitable for development phases~\cite{seward2004cachegrind, tao_detailed_2006}. To solve such issues, many sampling-based tools, such as perf~\cite{perf},
oprofile~\cite{oprofile}, are proposed to reduce the profiling overhead. They could attribute the percentage of cache misses to the specific lines of the source code based on the sampling. However, they cannot report both the \textbf{type} and \textbf{origin} of cache misses, making their reports not sufficient to guide the bug fixes.

Different types of cache misses, including compulsory misses, capacity misses, conflict misses, and coherency misses~\cite{1240655, jouppi1990reducing},  require different fixing methods, as detailed in Section~\ref{sec:overview}. A compulsory miss occurs when the related cache line is accessed for the first time, which is considered to be mandatory and unavoidable. Instead, a capacity miss may occur if the working set of a program exceeds the capacity of the cache. Capacity misses can be reduced by loop optimizations~\cite{loopoptimization} or array regrouping~\cite{ArrayTool}. In contrast, conflict misses can be introduced when more than N cache lines are mapping to the same set in N-way associative cache, and coherency misses may occur when multiple threads are accessing the same cache line simultaneously. Although some conflict and coherency misses can be reduced with padding, they require different padding strategies: fixing conflict misses should prevent the mapping to the same set, while coherency misses can be reduced by avoiding multiple threads accessing the same cache lines. Reducing cache misses also requires to know the origin: whether a problem is caused by the allocator or the application? For application bugs, which objects or which instructions are involved? Without knowing such information, it is impossible to reduce cache misses effectively.

Some tools aim to identify a specific type of cache misses, such as capacity misses~\cite{ArrayTool}, coherence misses~\cite{Cheetah, Laser, Huron, Feather}, and conflict misses~\cite{CCProf}. However, it is inconvenient to identify all types of cache misses using these tools, as they are designed as exclusive to each other. Further, none of them could correctly identify cache misses caused by the memory allocator. For instance, \texttt{cache-thrash} can be slowed down by $38\times$ when using \texttt{TCMalloc} (as shown in Table~\ref{tbl:effectiveness}). Without knowing the origin of cache misses, programmers may waste their efforts in improving the application but achieve only minor or no improvement. DProf~\cite{DProf} is the only tool that can identify all types of cache misses for data structures of Linux kernel, which unfortunately requires significant manual effort, as further discussed in Section~\ref{sec:related}.

This paper proposes a novel tool--\CP{}--that overcomes these shortcomings: (1) \CP{} is a unified profiler that could identify all fixable cache misses (except compulsory misses); (2) \CP{} only reports serious issues, saving manual effort spending on trivial issues; (3) \CP{} reports both type and origin of cache misses, providing useful information for bug fixes; (4) \CP{} only imposes reasonable overhead for its identification. Designing such a tool includes the following challenges.

\textit{The first challenge is to choose an appropriate profiling method that can classify all types of cache misses with reasonable overhead}. Prior work employs different sampling events, including address sampling for capacity misses~\cite{ArrayTool} and coherency misses~\cite{Cheetah,Feather}, HITM events~\cite{Laser} for coherency misses, and L1 cache misses for conflict misses~\cite{CCProf}. However, it is infeasible to combine these events together, as that will introduce prohibitive overhead and complexity. Although it is intuitive to employ the hardware-based sampling, the Performance Monitoring Units (PMU) supports up to hundreds of events (e.g., 207 events at Intel Xeon Silver 4114~\cite{pmutotalevents}). \CP{}'s selection is driven by the requirement of differentiating the type, reporting the origin, and measuring the seriousness of cache misses, as discussed later. In summary, such an event should capture the detailed information of memory accesses, such as the memory address, the related instruction, and the hit information (indicating a cache miss or not), which is often omitted by existing work~\cite{Cheetah, Feather}. Therefore, ``\textbf{the PMU-based precise address sampling}'' is chosen as the right event, and we elaborate why and how \CP{} exploits this event as follows.

\textit{The second challenge is to differentiate all
types of cache misses correctly.} The PMU-based sampling helps filter out cache misses, but it is impossible to correctly identify the type of each cache miss under the sampling, due to the lack of the history of cache usage and memory accesses. Instead, \CP{} proposes to identify coherency misses based on \textbf{the cumulative behavior of many misses}: only very few cache lines (not mapping to the same set) with extensive misses are most likely caused by coherency misses. Unfortunately, this rule cannot differentiate capacity misses from conflict misses, where the detailed access pattern is required for the differentiation, as further discussed in Section~\ref{sec:overview}. Further, CachePerf classifies other types of cache misses based on a key observation: \textit{serious cache
misses are typically caused by very few instructions whose access patterns are not altered during the
whole execution}. Based on this key observation, we propose a novel approach--\textbf{hybrid hardware
sampling}--to classify the type of cache misses: \textit{the PMU-based coarse-grained sampling detects
susceptible instructions with frequent cache misses, then the breakpoint-based fine-grained sampling
is employed to identify memory access patterns of these selected instructions.} This approach combines
the best of both worlds, as the coarse-grained sampling could reduce the profiling overhead, while
the fine-grained sampling collects a short history of memory accesses that is necessary to classify
the access pattern. For instance, it is easy to determine conflict misses if multiple continuous accesses are accessing the same cache set.

\textit{The third challenge is to differentiate cache misses caused by the memory allocator from those ones caused by applications}. Although allocator-induced cache misses may have a high impact on the performance, they get less attention than they deserve. This paper makes the following \textit{observations}: (1) the allocator may introduce both conflict and coherence misses (mainly false sharing, a type of coherency misses that multiple threads are accessing different words of the same cache line~\cite{Sheriff}); (2) Allocator-induced cache misses share the same attribute that multiple heap objects are involved unnecessarily, although this is not the sufficient condition. For instance, allocator-induced false sharing should have more than two objects on the same cache line. Further, these objects, accessed by different threads, must be allocated by different threads.  Similarly, an allocator may introduce conflict misses, when multiple objects are mapped to the same set of cache lines. To the best of our knowledge, \textit{\CP{} is the first work that reports allocator-induced cache misses}.

\CP{} further designs practical mechanisms that help reduce the detection overhead and avoid reporting minor issues:  (1) \CP{} tracks a specified number of the most recent memory accesses (or a window), and then only checks cache misses inside if the miss ratio (i.e., the number of misses divided by the number of accesses) in the current buffer is larger than a
threshold. This windowing mechanism also helps filter out sporadic cache misses, e.g., compulsory misses; (2) \CP{} further proposes a ``ratio-based filtering'' that only reports an issue if the ratio of memory accesses or cache misses is larger than a threshold; 

%Existing tools may report trivial issues due to their imperfect metrics~\cite{Sheriff, Cheetah, Feather, CCProf}. For instance, some utilize the same absolute metric for different applications~\cite{Sheriff, Cheetah, Feather}, e.g., the number of cache invalidations to evaluate false sharing issues. However, the same number of cache invalidations may have different performance impacts on a long-running or short-running program. Instead, \CP{} proposes an application-specific metric---``summarized access ratio''---that considers the potential impact of the related instructions on each application. \CP{} further proposes a windowing mechanism to filter out sparse cache misses, e.g., compulsory misses, as discussed in Section~\ref{}.

We evaluated \CP{} on a range of well-studied benchmarks and real applications, where some have known cache misses. Based on our evaluation, \CP{} only introduces 14\% performance overhead and 19\% memory overhead (for large applications), while detecting all known bugs and night previously-unknown cache misses. Guided by \CP{}'s report, we are able to fix most detected cache misses, achieving the performance speedup up to $38\times$. Overall, the paper
makes the following contributions:

\begin{itemize}
    \item It proposes a novel hybrid sampling scheme that combines coarse-grained PMU-based sampling and fine-grained breakpoint-based sampling, with a better trade-off between performance and accuracy.  
    
    \item It is the first tool that can classify different types of cache misses without manual involvement. 
    
    \item It proposes practical mechanisms to differentiate cache misses caused by the allocator from those from applications, and to prune insignificant issues.
    
    \item It provides the detailed implementation of a  profiler with low overhead (14\% on average) and high effectiveness, confirmed by our extensive evaluation.
    
\end{itemize}

\section{Background and Overview}
\label{sec:overview}

This section first introduces some basic background of cache misses, and then discusses the basic idea of \CP{}. 

\subsection{Types of Cache Misses}
\label{sec:misstype}

Cache miss can be classified into compulsory miss, capacity miss, conflict miss, and coherence
miss~\cite{DProf}. Among them, a compulsory miss occurs when the cache line is accessed for the first time,
which is mandatory and unavoidable~\cite{jouppi1990reducing}. In the remainder of this paper, we mainly focus on the
other three types of cache misses. In the following, we will discuss their definitions, fix strategies,
and possible causes.

\subsubsection{Capacity Miss} 

\begin{figure*}[!h]
\centering
\includegraphics[width=4.5in]{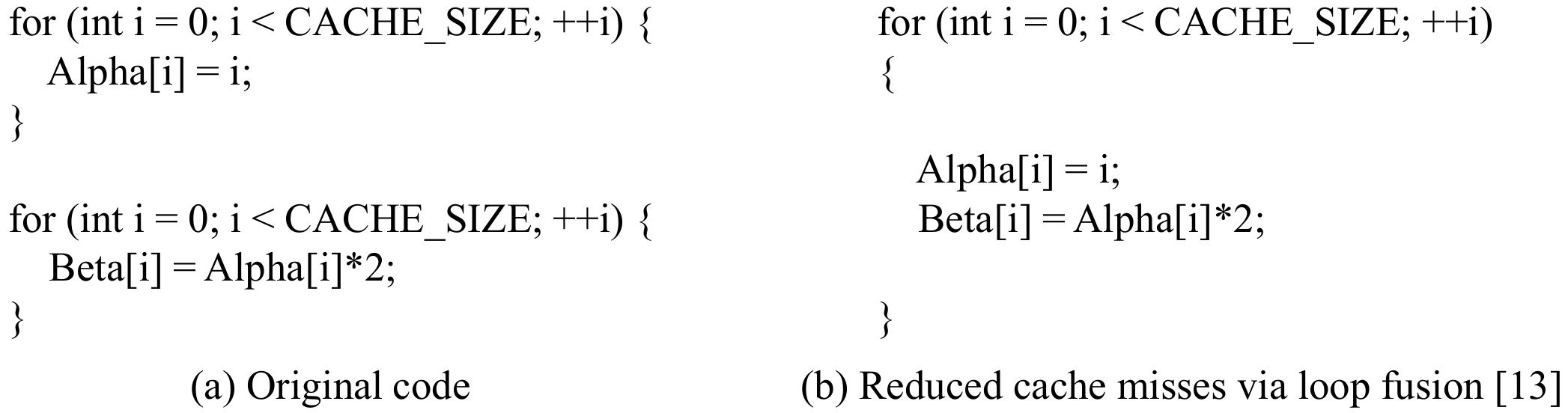}
\captionsetup{justification=justified}
\caption{An example with capacity misses. 
% \todo{the title of (b) can be changed to ``Reduced Cache Miss via Loop Fusion''}
\label{code:capacity} 
}
\end{figure*}

Capacity misses occur when the accessed data of a program exceeds the capacity of the cache~\cite{capacitymiss}. When the cache cannot hold all the active data, some recently-accessed cache lines are forced to be evicted, which leads to cache misses if they are accessed again.
As shown in Fig.~\ref{code:capacity}(a), both \texttt{for} loops will
suffer from cache capacity misses, as both \texttt{Alpha} and \texttt{Beta}’s size is four times of the cache size
(with the “CACHE\_SIZE” number of integers).

%\todo{The severeness of capacity misses will be greater when applications are consuming more data. } // This is not precise
Capacity misses are mainly caused by applications. Not all capacity misses can be completely eliminated. However, some can be significantly reduced via array regrouping~\cite{ArrayTool} or loop optimizations~\cite{loopoptimization} (e.g., loop tiling~\cite{looptiling}).  For the example shown in Fig.~\ref{code:capacity}(a), we could combine two loops into one loop as Fig.~\ref{code:capacity}(b) to reduce cache misses, also known as loop fusion~\cite{loopfusion}.

\begin{figure*}[!ht]
\centering
\includegraphics[width=5.4in]{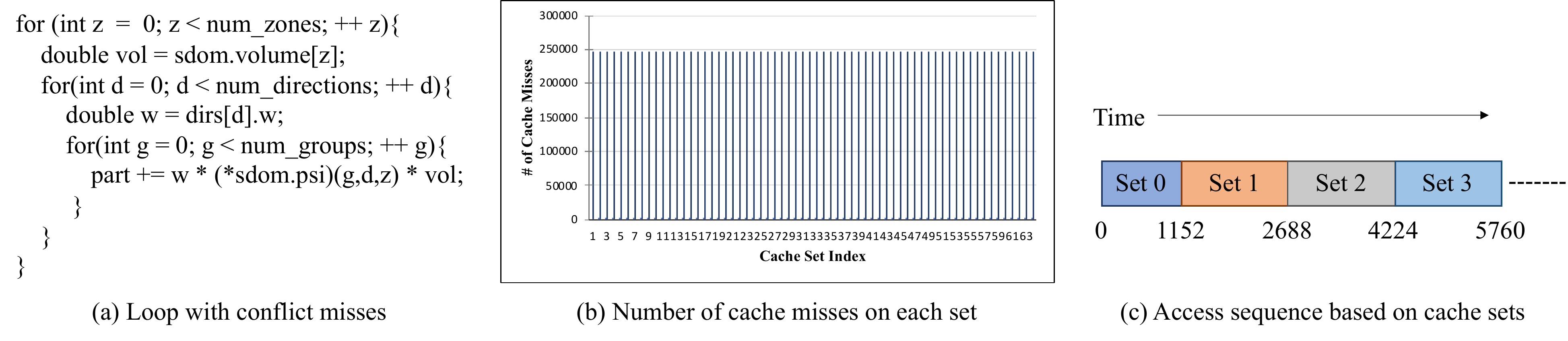}
\caption{A real example of conflict misses from the \texttt{Kripke} application~\cite{Kripke, CCProf}.
\label{fig:conflict}}
\end{figure*}

\subsubsection{Conflict Miss} 
Conflict misses are introduced in direct-mapped or set-associative cache~\cite{CCProf, DProf}. 
For an N-way associative cache, conflict misses will occur when more than N cache lines mapping to the same set are accessed recently. 

Fig.~\ref{fig:conflict}(a) shows a real example of conflict misses:  \texttt{Kripke} accesses multiple cache lines of Set 0 and then Set 1, as shown in Figure~\ref{fig:conflict}(c). For this example, each cache set has exactly the same number of cache misses, as shown in Figure~\ref{fig:conflict}(b). This indicates that conflict misses cannot be identified by the portion of misses in cache sets. Instead, the access pattern of the corresponding instruction(s) should be employed to identify such issues. 

Conflict misses can not only be caused by applications, but also can be caused by the allocator when multiple allocated objects are mapped to the same cache set. \texttt{raytrace}, an application in PARSEC~\cite{PARSEC}, introduces a 27\% performance slowdown due to conflict misses of the allocator, as shown in Table~\ref{tbl:effectiveness}. Conflict misses can be resolved or reduced by changing the starting addresses of objects,  or padding the corresponding structure. Even for allocator-induced applications, we could insert some bogus memory allocations or change the size of the corresponding allocations in order to reduce cache misses. 

\subsubsection{Cache Coherence Misses}
Multithreaded applications are prone to coherence misses when multiple threads are accessing the same cache line. When a thread writes to a cache line, the cache coherence protocol invalidates all existing copies of this cache line, introducing \textit{cache coherency misses}. Coherence misses can be caused by true and false sharing. False sharing occurs when multiple threads are accessing different portions of the same cache line, while threads are accessing the same units in true sharing. When modern architectures are equipped with larger cache lines and more cores, they are more prone to coherence misses with higher performance impacts. 

True sharing is typically caused by applications.
Although true sharing is considered to be unavoidable~\cite{Sheriff}, programmers could still refactor the code to reduce its seriousness~\cite{Laser, MySQLBug}. For example, programmers may reduce the updating of shared variables by using thread-local or local variables. False sharing can be caused by both applications and allocators. For allocator-induced false sharing, multiple threads may access different objects concurrently within the same cache line that are allocated by different threads. False sharing can be reduced by padding the data structure~\cite{Huron}, or using per-thread private pages~\cite{Sheriff}. Therefore, it is important to differentiate between false and true sharing, as they need different fixing strategies.

\subsection{Basic Idea of \CP{}}
% \label{sec:differentiation}

\CP{} aims to identify the type and the origin of cache misses correctly so that programmers can further fix them correspondingly. More specifically, \CP{} not only differentiates capacity misses, conflict misses, and coherency misses, but also differentiates whether some misses are caused by the allocator or the application. If they are caused by the application, \CP{} further reports the lines of code with the issue, e.g., call sites and instructions. For allocator-induced cache misses, \CP{} also reports the sizes of the related objects. 

\subsubsection{Differentiating Different Types of Misses}
\label{sec:differentiation}

As mentioned in Section~\ref{sec:intro}, it is challenging to identify the type of each miss directly. For instance,
to identify a capacity miss, it is required to know the working set of the current program~\cite{DProf}, which
is infeasible under the coarse-grained sampling. CCProf~\cite{CCProf} observes that ``a relatively larger portion of cache misses in a subgroup of the total
cache sets over the others indicates conflicts in those cache sets''. Unfortunately, this method is
\textit{neither sufficient nor necessary} condition of conflict misses, although it seems to be valid at the first glance. As shown in Fig.~\ref{fig:conflict}(b), all cache sets have exactly the same number of cache misses for the \texttt{Kripke} application. However, this issue belongs to ``conflict misses'' based on the access pattern shown in Fig.~\ref{fig:conflict}(c). Further, a \texttt{for} loop consecutively accessing an array (e.g., 1.5 times larger than the cache size) may cause only half of the cache sets to have significantly more cache misses than the other half, but this belongs to capacity misses instead of conflict misses.

In fact, \CP{}'s identification is based on the following observations: (\rom{1}) Coherence misses typically occur on few cache lines, but not for capacity and conflict misses; (\rom{2}) 
Extensive cache misses are typically caused by few susceptible instructions; (\rom{3}) The patterns of memory accesses are necessary to differentiate conflict misses from capacity misses: if multiple memory accesses are accessing the same set of cache lines, then it is an issue of conflict miss; If they are accessing different cache sets, the issue is more likely to be capacity miss.

\textit{Observation (\rom{1})} indicates that coherence misses (e.g., false sharing and true sharing) can be identified by checking the cumulative behavior of cache lines: \textit{if few cache lines (not on the same set) have more cache misses than others, then this issue must be caused by coherence misses.} Like existing work~\cite{Sheriff}, false sharing can be easily differentiated from true sharing using their definitions: if multiple threads are accessing different words of the same cache line, then it is false sharing. Otherwise, it is true sharing. We will use \textit{the Performance Monitoring Unit (PMU)'s address sampling} to collect accesses on a cache line, helping differentiate false sharing from true sharing. 

Based on \textit{observation (\rom{1}) \ and (\rom{2}}), we propose the \textbf{hybrid hardware sampling} to classify cache misses: \textit{the hardware Performance Monitoring Unit (PMU) is employed to collect the coarse-grained samples in order to pinpoint susceptible instructions with extensive cache misses; After that, the breakpoints are further installed on these instructions in order to collect fine-grained memory accesses to understand their memory access patterns.} After collecting memory access patterns, it is possible to differentiate conflict misses from capacity misses using \textit{observation (\rom{3})}.  

\subsubsection{Differentiating Serious Issues from Minor Ones}
Minor issues, although they are not false
positives, should be excluded to avoid wasting the time of programmers. Unfortunately, most
existing tools~\cite{ArrayTool, Sheriff, Predator, Laser, Cheetah, CCProf, Feather} cannot achieve this goal, as they typically utilize the same absolute metric for different applications, e.g., the number of cache invalidations to evaluate false sharing issues, omitting the temporal effect. However, the same number of cache misses may have different performance impacts for a long-running or short-running program. Further,
a program with sparse cache misses and another one with intense misses may benefit differently from the reduction of cache misses, even if they have a similar execution length and cache misses.

\CP{} further proposes two ratio-based mechanisms to exclude minor issues. First, \CP{} proposes a \textit{windowing} mechanism that tracks a specified number of the most recent memory accesses, and then only checks cache misses inside if the miss ratio (i.e., the number of misses divided by the number of accesses) in the past window is larger than a threshold, as discussed in Section~\ref{sec:missRatioChecker}. This windowing mechanism  excludes sparse or sporadic cache misses.   
Second, \CP{} only reports a potential issue if its related  memory accesses and cache misses are higher than 0.01\% and 1\% separately. The access ratio can be utilized to predict the potential performance impact. Assuming that the memory access is  $200\times$ slower than the L1 cache access~\cite{misspenalty}, and  the access ratio of a potential bug is 0.01\% of the total accesses (of the program). We further assume that other accesses of this program can be satisfied at L1 cache, then the total runtime of this program is \mbox{0.01\% $\times$ 200X + 99.99\% $\times$ X = 101.9\%X}, if the cycle of L1 cache access is $X$. Then this bug will introduce at most  2\% slowdown, comparing to all accesses are satisfied by the L1 cache ($100\%X$). Similarly, the ratio of cache misses helps prune insignificant instructions.

\subsubsection{Differentiating Allocator-Caused Misses from Applications.}
\label{sec:differentiateAllocator}
As discussed in Section~\ref{sec:misstype}, the
allocator may introduce both conflict and coherence misses. When an allocator allocates multiple 
objects that happen to access  few sets of cache lines, it introduces conflict misses. An allocator can introduce false sharing by allocating multiple objects in the same cache line to different threads~\cite{Hoard}. CachePerf tracks the allocation information (e.g., the thread, address) that could help differentiate the bugs caused by applications from those caused by the allocator. To the best of our knowledge, CachePerf is the first work that could report allocator-induced cache misses.

\section{Design and Implementation}
\label{sec:implement}

This section discusses the detailed design and implementation of \CP{}. \CP{} is designed as a library that can be linked with different applications, without the need of changing and recompiling user programs. In the following, we start with the description of \CP{}'s basic components, and then discuss each component separately. 

%\subsection{Basic Components of CachePerf}
\label{sec:basicidea}
\begin{figure*}[!ht]
\centering
\includegraphics[width=5.4in]{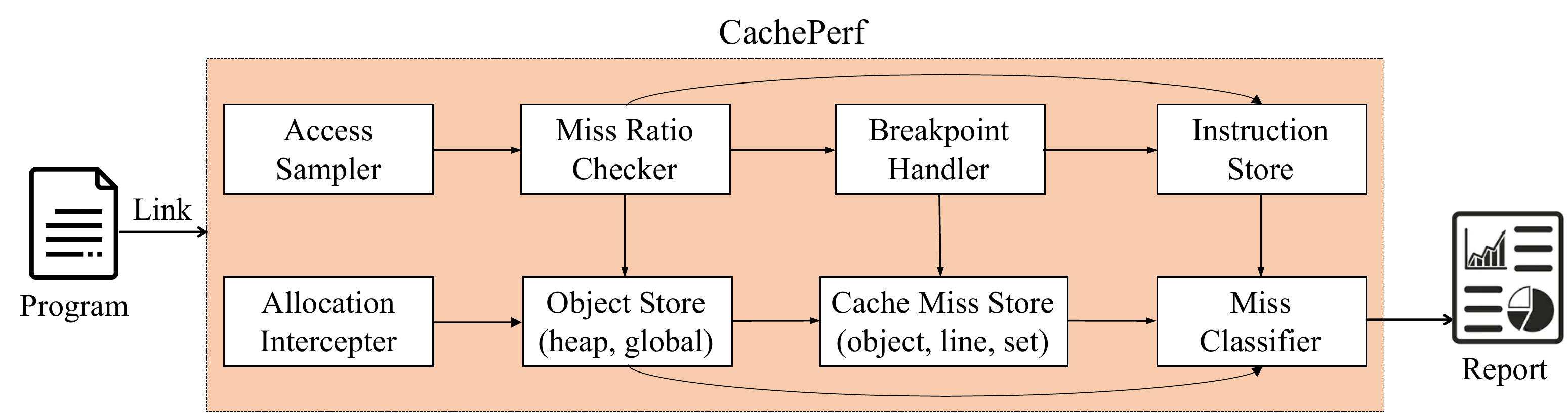}
\caption{Basic components of \CP{} \label{fig:framework}}
\end{figure*}

Fig.~\ref{fig:framework} shows the basic components of \CP{}. As mentioned in Section~\ref{sec:intro}, \CP{} relies on the PMU-based sampling to collect the information of memory accesses and cache misses, which will be handled by its ``\textit{Access Sampler}'' module. To exclude insignificant cache misses,  \CP{} introduces a ``\textit{Miss Ratio Checker}'' module that computes and checks the cache miss ratio (the percentage of cache misses in all memory accesses). When the cache miss ratio is larger than a predefined threshold (e.g., 0.5\%), as further discussed in Section~\ref{sec:thresholds}, all recent cache misses will be further updated to ``\textit{Miss Store}'' and ``\textit{Instruction Store}''. Otherwise, all cache misses will be skipped. Due to this filtering mechanism, low-frequency cache misses (such as some compulsory misses) will be excluded automatically. When continuous cache misses from the same instruction are detected or multiple misses are landing on the same cache set, indicating possible capacity or conflict misses, \CP{} further employs breakpoints to collect fine-grained memory accesses information (via ``breakpoint Handler''), which enables us to differentiate conflict misses from capacity misses.
%\JZ{or: and differentiates conflict misses from capacity misses. } 
% Figure \ref{fig:framework} shows the basic components of \CP{}.

In order to attribute cache misses to data objects (called ``data-centric'' analysis~\cite{ArrayTool}), \CP{} further intercepts memory allocations and deallocations, and updates the ``\textit{Object Store}'' correspondingly. ``Object Store'' tracks address ranges and callsites of heap objects. In the end, \CP{} classifies cache misses by integrating the data in ``\textit{Miss Store}'' and ``\textit{Instruction Store}'', and finally reports helpful information based on ``\textit{Object Store}'', including the allocation call sites, object size, and object name (only for global objects).  
Different from existing tools~\cite{CCProf, Feather}, there is no need for offline analysis, i.e., it has no hiding overhead. 
%\JZ{or: \CP{} does not depend on any offline data processing or visualization.}
 
\subsection{Access Sampler} 

For the access sampler, \CP{} employs the Performance Monitoring Units (PMU) to sample memory accesses. The PMU is the ubiquitous hardware in modern architectures (e.g., X86 or ARM) that can provide hundreds of hardware events~\cite{intelmsr}. There is a trend for profilers to build on top of the PMU~\cite{DBLP:conf/sc/ItzkowitzWAK03, DBLP:conf/sc/BuckH04, ibs-sc, ibs-pact, Sheng:2011:RLN:1985793.1985848, Jung:2014:AML:2568225.2568311, Laser, Cheetah}, due to its low overhead. Currently, Linux also provides a system call --\texttt{perf\_event\_open} -- that allows to configure and start the PMU easily. 

\begin{table}[!ht]
\centering
% \scalebox{0.8}{
\begin{tabular}{c|c|c}
\hline
Configuration                 & Load Sampling         & Store Sampling       \\ \hline
type                          & \multicolumn{2}{c}{\texttt{PERF\_TYPE\_RAW}}         \\ \hline
config                        & 0x1cd                 & 0x82d0               \\ \hline
sample\_period                & 20000 ($\pm10\%$)        & 50000 ($\pm10\%$)        \\ \hline
freq                          & \multicolumn{2}{c}{false}                   \\ \hline
\multirow{3}{*}{sample\_type} & \multicolumn{2}{c}{\texttt{PERF\_SAMPLE\_IP |}}      \\
                              & \multicolumn{2}{c}{\texttt{PERF\_SAMPLE\_ADDR} |}    \\
                              & \multicolumn{2}{c}{\texttt{PERF\_SAMPLE\_DATA\_SRC}} \\ \hline
precise\_ip                   & 3                     & 1                    \\ \hline
\_\_reserved\_1               & 3                   & 0                    \\ \hline
config1                       & 3                     & 0                    \\ \hline
\end{tabular}
% }
\caption{Configuration of the PMU sampling\label{tbl:sampling}}
\end{table}

\CP{} samples two types of events, including memory loads and stores. The configuration for the PMU sampling is shown in Table~\ref{tbl:sampling}, which is based on Intel's Xeon machine. 
To balance the detection effect on loads and stores, we empirically set the sampling period of loads as 20,000, and the one of stores as 50,000, which has been evaluated in Section~\ref{sec:thresholdImpact}. To avoid different threads from sampling the same instructions, we introduce 10\% randomized variance for each thread's sampling period. 
Note that it is important to include \texttt{PERF\_SAMPLE\_DATA\_SRC} in the sample type so that we can know which level the corresponding instruction is hit, such as L1, L2, LLC, or memory. It is also referred to as ``hit information'' in the remainder of this paper. 
%For memory loads, it is important to set the \texttt{config1} to be 3 as the latency threshold \JZ{to avoid sampling bias. latency larger than 3. 3 is the minimum number of loading events. }. 
%\HM{1.which is also called->is also referred to as 2. Why setting config1 to 3? Maybe it's better to give a reason when you set a configuration}

\CP{} employs the following information of the sampling:  the type of access (e.g., load or store),  hit information, memory address, and instruction pointer (IP). Among them, the hit information helps identify all cache misses from all sampled memory accesses, where all accesses that do not hit on the L1 cache will be treated as cache misses. IP tells the instruction performing the corresponding access, and the memory address helps pinpoint which cache line and cache set have the miss, enabling us to perform the classification. 

\begin{comment}
\begin{table}[!ht]
    \centering
    \begin{tabular}{l|l|l}
        Information & Description & Used\\ \hline
        ID & Sampling Sequential Number & \\
         Access Type & Load or Store Access & \checkmark \\
         PID and TID & Process ID and Thread ID  &\\
         Timestamp & Sampling Timestamp & \\
         Address & Memory Access Address & \checkmark\\ 
         Level & Hit Information & \checkmark\\ 
         IP & Instruction Pointer & \checkmark\\ 
         CPU & CPU Index \\
         Latency & Memory Access Latency \\ \hline
    \end{tabular}
    \caption{Information of Each Access Miss \label{tab:accessinfo}}
    \vspace{-0.2in}
\end{table}

\end{comment}

\subsection{Miss Ratio Checker} 
\label{sec:missRatioChecker}
%\TP{Windowing mechanism to filter out sporadic cache misses, as serious cache misses will typically have very high cache miss ratio}. 
A miss ratio checker is introduced to filter out sparse cache misses. As mentioned above, since sparse cache misses may not incur significant performance slowdown, they should be excluded in order to avoid wasting the effort of fixing such issues. Further, the filtering reduces the memory overhead of storing such cache misses and the performance overhead of spending in classification.

In the implementation, \CP{} maintains two circular buffers to track the most recently sampled memory accesses for each thread, one buffer for memory loads and the other one for memory stores. 
These buffers are updated in First-In-First-Out order that the later accesses will overwrite the least-recent memory accesses. \CP{} computes the cache miss ratio upon every access via dividing the number of cache misses by that of accesses. Only when the miss ratio of the buffer is larger than a predefined threshold (e.g., 0.5\%), all cache misses in the current buffer will be handled and be updated to ``\textit{Instruction Store}'' and ``\textit{Miss Store}''. Otherwise, they will be skipped. The Instruction Store holds the information related to instructions, such as the number of accesses and cache misses. The Miss Store maintains the detailed information about each cache miss, e.g., object, line, and set. 
%\HM{Miss ratio checker is introduced to achieve this target->To achieve this, a miss rate checker is introduced}

\subsection{Breakpoint Handler} 
\label{sec:breakpointHandler}
As mentioned in Section~\ref{sec:overview}, \CP{} employs the breakpoints to collect fine-grained memory accesses of the selected instructions, enabling us to differentiate conflict misses from capacity misses. For the susceptible instructions, \CP{} focuses on two types of instructions:
(1) instructions introducing multiple continuous cache misses, indicating that they may incur extensive cache misses. (2) instructions introduce extensive misses on the same set in a time window, which are potential candidates for conflict misses. 

After identifying these instructions, \CP{} installs hardware breakpoints via the \texttt{perf\_event\_} \texttt{open} system call by specifying the \texttt{type} to be  \texttt{``PERF\_TYPE\_BREAKPOINT"}  and the \texttt{bp\_type} to be \texttt{``HW\_BREAKPOINT\_X''}. 
After the installation, every time a program executes such an instruction, \CP{} will be interrupted so that it could collect the fine-grained memory accesses of each instruction. However, the interrupt handler provides no information about the memory address, as the breakpoint is typically triggered before the access. \CP{} infers the memory address by analyzing the corresponding instruction. For example, if the instruction is ``\texttt{addl  \$0x1,-0x4(\%rbp)}'', then \CP{} could infer the stored memory address via the value of register and \texttt{rbp}. \CP{} employs Intel's \texttt{xed} library~\cite{xed} to perform the binary analysis. 
%That is, it will analyze the binary instruction to compute the particular memory address.     

% How many instructions could we support. 
To simplify the handling, \CP{} only installs one breakpoint for all threads at a time, collecting all accesses from different threads. To reduce the overhead caused by handling endless interrupts, \CP{} only collects at most 64 accesses from one instruction. If there are 8 accesses landing on the same set, then it is identified as a bug with conflict misses. Otherwise, it is a bug with capacity misses. Based on this, if 8 continuous accesses are landing on the same cache set, which can be clearly identified as ``conflict misses'', \CP{} will remove the breakpoint so that it could monitor other instructions. 

However, it is possible that an instruction has no or few accesses after the installation. When new instructions require to be monitored, \CP{} further introduces an expiration mechanism that a breakpoint will be expired after 100ms. In this way, \CP{} is able to install breakpoints on new instructions. The identification of cache misses is further discussed in Section~\ref{sec:classifier}. 
%\HM{add reference section}

%If a cache set is touched more than 8 times in the next 64, \CP{} will view the instruction as a cause of cache conflict misses, and will report it at the end of the program.

% What is the handling? Should we avoid race conditions? 

% How many interrupts should we use? 

\subsection{Important Data Stores}
\CP{} maintains \textit{Object Store}, \textit{Miss Store}, and \textit{Instruction Store}, as further described below. 

% \todo{What type of data structure? How they will be updated?} 
\paragraph{Object Store} 
Object Store tracks the information of two types of objects, heap objects and global objects, as most cache misses occur on these objects, 
%Instead, stack variables are typically never be the target of extensive cache misses.  
which are handled differently.  

For heap objects, \CP{} intercepts all memory management functions, such as \texttt{malloc()} and \texttt{free()}, in order to track their corresponding callsites. For each heap object, \CP{} tracks its size, callsite, and address range. As there are large amounts of heap objects, the Data Store should be carefully designed in order to support the following operations efficiently: adding and updating of an object via the starting address upon memory allocations and deallocations, and searching by an address in the range of a valid object upon each sampled access. Although the hash table can support the adding and updating operations efficiently via the starting address (as the key), it is expensive to search the memory address inside heap objects (different from the key). Instead, an ordered list/array supports the searching better via the binary search. Furthermore, we observe that heap objects are typically classified into small, medium, and large sizes, where the number of small-size objects is much larger than that of medium-size and large-size objects.  

\begin{figure}
    \centering
    \includegraphics[width = 3.25in]{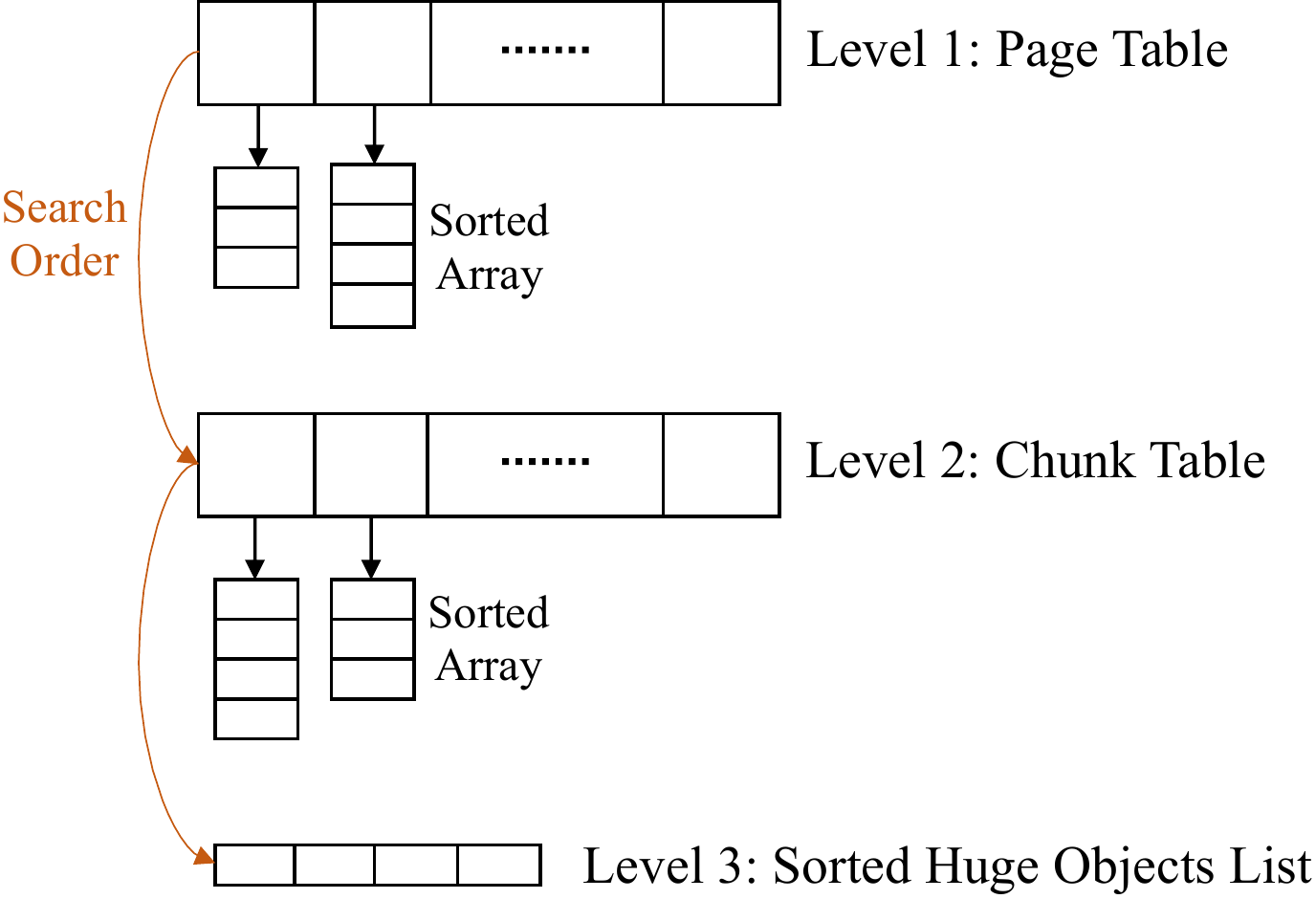}
    \captionsetup{justification=justified}
    \caption{A three-level object store that combines with the shared memory and the sorted array/list. }
    \label{fig:3leveltable}
\end{figure}

Based on these observations, \CP{} designs a three-level data store as shown in Fig.~\ref{fig:3leveltable} to support efficient adding, updating, and searching. In particular, any object can be stored in one of Page Table, Chunk Table, and Sorted Huge Objects List, which are mutually exclusive. 
%In particular, every virtual page has an entry in the page table, which has a sorted link list that lists all objects on this page. Similarly, every virtual chunk (e.g., one MB) has one entry in the chunk table. \CP{} reserves a large chunk of memory (continuously) for Page Table and Chunk Table so that it could support fast lookup. 
\CP{} updates these tables/lists as follows: (1) if an object exists only in a single page, it is stored in the \textit{Page Table}; (2) If the range of an object crosses two different pages but within the same megabyte, it is stored in the \textit{Chunk Table}; (3) Otherwise, it is inserted into the \textit{Sorted Huge Objects List}.
For each address, \CP{} always searches the Page Table at first (with the highest possibility), then the Chunk Table, and finally the Sorted Huge Objects List, and stops if the object is found already. Although three searches are required for some objects, however, we believe that such searches will be fewer than others. This design is based on the assumption that small objects (less than 1-page size) are typically significantly more than large objects. 

For the performance reason, each entry of \textit{Page Table} and \textit{Chunk Table} stores a pointer pointing to a sorted array that stores all allocated objects inside the same page (4KB) and chunk (1MB). If an entry is empty (with the NULL value), then there are no objects in the corresponding page or chunk, indicating the unnecessary of searching for a higher-level store. Both tables are employing the shadow memory~\cite{Zhaoqin, AddressSanitizer} to store these pointers, where the index of each entry could be computed simply with a bit-shifting operation. Since the number of huge objects is typically small, a sorted list is used to store huge objects, which is not using the shadow memory.  For the sorted arrays and lists, the search can be done efficiently via the binary search.

\CP{} also proposes a callsite-based optimization to reduce the overhead, especially on the updates of memory allocations and sampled accesses: if objects from a particular callsite have a much lower cache miss ratio, compared to the average one, then all allocations and  cache misses from this callsite could be safely skipped. 
%\TP{Jin:  the percentage of allocation size as the metric. For instance, if this callsite allocates about 2\% objects by size, but all misses from this callsite is less than 2\%.}
Based on our observation, such an optimization reduces the overhead by over 30\% for a particular application (\texttt{Canneal} of PARSEC~\cite{PARSEC}).  

%: Since objects with cache misses are usually allocated in few specific callsites, we can only add and update metadata of objects from those problematic callsites, after the heap object table has saved information for a number of objects. canneal improves the performance by 30\% 40\%. If less than 1\% of miss ratio, the misses less than the overall misses. 

\CP{} handles global objects differently, as an application typically has a small number of global objects and they are not increased during the execution. All accesses of global objects (e.g., addresses) will be stored in a hash table. \CP{} obtains the name and address range by analyzing the corresponding ELF header, and then computes the miss ratio of each object, as discussed in Section~\ref{sec:classifier}. 
%metadata before doing classification. It reads the ELF header of the program to get global object data and saves them as an array sorted by object starting addresses. 

%When \CP{} reports the global object metadata, it will do range searching in the Object Store to find the corresponding metadata.

\paragraph{Miss Store} 
The Miss Store saves the information of each cache miss, which is not filtered out as described by Miss Ratio Checker (Section~\ref{sec:missRatioChecker}). In particular, cache misses are stored in two separate data structures:  an array (with the size of the number of cache sets) stores the information of each cache set, and a hash table stores the information of each cache line (using the starting address as the key). For both data structures, \CP{} stores the number of cache misses (on each cache set and each cache line). For cache lines, \CP{} further stores the thread information for accessing each word, which could help differentiate false sharing from true sharing.

%For each cache set, \CP{} records the number of cache misses. To identify the misses are introduced by the allocator or the application, it tracks the number of heap objects with issues. To provide helper information, it saves the sizes and addresses of heap and global objects with issues and also keeps allocation callsite of heap objects by a hash table.

%For each cacheline with cache misses, \CP{} saves the same types of metadata as cache sets. In order to classify cache coherency misses, \CP{} also records thread indexes for each word, to record threads that have stored into the word.

\paragraph{Instruction Store}
 Instruction Store saves the information of memory accesses (e.g., loads and stores) and cache misses of the selected instructions by the miss ratio checker (as discussed in Section~\ref{sec:missRatioChecker}). The data structure of Instruction Store is a hash table that uses the instruction pointer as the hash key. For each instruction, \CP{} records the number of cache misses, the related cache set, and the detailed memory access pattern. 
 
Since each line/statement of the code may be related with multiple instructions (at the assembly level), \CP{} further summarizes cache misses of the related statement, and only reports statements with extensive cache misses. %\todo{To achieve this, \CP{} utilizes two hash tables: the first level hash table stores the mapping between the instruction pointer and the statement;  the second level hash table utilizes the statement as the key, and then stores the information of each statement.} 

%Multiple IPs may relate to the same statement. (IP -- key, data --> addr2line ) Two levels of hash table: one is to use IP to look for the code (addr2line). Using the key to find the hashcode. 

\subsection{Miss Classifier}
\label{sec:classifier}

\RestyleAlgo{ruled}
\SetKwComment{Comment}{/* }{ */}

\begin{algorithm}[h]
\caption{The Algorithm of Classifying Cache Misses\label{alg:classifier}}

\For{cache line \texttt{c} in Miss Store}
{
\If{multiple threads access the same words of \texttt{c}}
{
Report \texttt{true sharing}
}
\If{multiple threads access  different words of \texttt{c}}
{
\eIf{\texttt{c} has multiple objects allocated by different threads}
{
Report \texttt{allocator-induced false sharing} 
}
{
Report \texttt{application's false sharing}
}
}
}

\For{instruction \texttt{i} in Instruction Store}
{
\If{the issue is reported as \texttt{coherency miss}}
{
\textbf{continue}
}
\eIf{\texttt{i}'s misses land on the same cache set}
{
\eIf{misses are landing on multiple heap objects}
{
Report \texttt{allocator-induced conflict miss}
}
{
Report \texttt{application's conflict miss}
}
}
{
Report \texttt{application's capacity miss}
}
}
\end{algorithm}

\CP{} classifies and reports serious cache misses by combining the information from Cache Miss Store and Instruction Store together. The detailed algorithm is shown as Algorithm~\ref{alg:classifier}. \CP{} omits cache misses without significant performance impacts. Instead, it focuses on instructions or cache lines that have passed the ``ratio-based filtering'': (1) for an application, if the number of load misses is less than 3\% of all load accesses and the number of store misses is less than 1\% of all store accesses, then \CP{} will not report any issue; (2) for each instruction, if its memory accesses are less than 0.01\% of total accesses, or its cache misses are less than 1\% of total misses, this instruction will not be reported; (3) for each cache line and each cache set, it will be reported only if its misses larger than 1\% of all misses. These numbers are set based on our experience, which has been evaluated as Section~\ref{sec:thresholdImpact}.

\CP{} reports potential coherence misses by checking all cache lines in Miss Store. As discussed in Section~\ref{sec:differentiation}, few cache lines with extensive cache misses but not mapping to the same cache set can be caused by coherency misses.  For each cache line, \CP{} can further determine the type, false sharing and true sharing, via word-level information of the corresponding cache lines. If multiple threads are accessing the same words of the cache line, then it is true sharing of applications. Otherwise, it is a  false sharing problem. \CP{} further checks whether multiple objects on the same cache line are allocated by different threads or not. If yes, then it is allocator-induced false sharing. Otherwise, it is the application's false sharing. If cache lines are identified as coherency misses, the corresponding instructions will be marked as checked, which will be excluded for identifying conflict misses and capacity misses afterward. 

\CP{} differentiates capacity misses from conflict misses based on the memory access pattern of each instruction (with extensive cache misses) in the Instruction Store. A simple mechanism is employed to differentiate conflict misses from capacity misses: if the number of accesses mapping to the same cache set is larger than a threshold (e.g., 8), then the corresponding cache misses will be considered as conflict misses. Otherwise, they are capacity misses. 
For conflict misses, \CP{} further checks whether they are caused by the allocator or not: if they are involved with multiple heap objects, this belongs to allocator-induced conflict miss. Otherwise, it is an application's conflict miss. 

\CP{} could further report the detailed information of cache misses, including the instruction information (from Instruction Store) and object information (from Object Store). The former one tells which instructions introduce cache misses, while the latter one helps locate the heap object with its allocation callsite. This information could guide bug fixes. For instance, if two objects mapping to the same cache set introduce excessive cache misses, such an issue can be significantly reduced by changing the address of objects (by mapping to different sets). 

\section{Experimental Evaluation}
\label{sec:evaluation}

The experimental evaluation will answer the following research questions:
\begin{itemize}
\item How is the effectiveness of \CP{}? (Section~\ref{sec:effectiveness}) 	
\item What is the performance overhead of \CP{}? (Section~\ref{sec:perf})
\item What is the memory overhead of \CP{}? (Section~\ref{sec:memory})
\item What are the impacts of different configurations? 
(Section~\ref{sec:thresholdImpact})
\end{itemize}

\subsection{Experimental Setting}
\label{sec:environment}

\paragraph{Hardware Platform:} Experiments are evaluated on a two-processor machine, where both processors are Intel(R) Xeon(R) Gold 6230 with 20 cores. We only enabled 16 hardware cores in one node to exclude the NUMA impact as it is outside the scope of this paper. The machine has 256GB of main memory, 64KB L1 cache, and 1MB of L2 cache. 

\paragraph{Software:} The OS is Ubuntu 18.04.3 LTS, installed with Linux-5.3.0-40. The compiler is GCC-7.5.0, while we are using \texttt{-O2} and \texttt{-g} flags for the compilation.

\paragraph{Evaluated Applications:} 
Two types of applications are included in the evaluation, including general applications and applications known to have cache misses. 
%Buggy applications were collected from the following sources: conflict misses~\cite{CCProf}, capacity misses~\cite{ArrayTool}, and coherency misses~\cite{Feather}. 
In particular, all 13 applications from the PARSEC-2.0 benchmark are included as general applications~\cite{PARSEC}, but some also have known bugs. Buggy applications with coherence misses (false sharing) include two stress tests \texttt{cache-scratch} and \texttt{cache-thrash} from Hoard~\cite{Hoard}, and two Phoenix~\cite{Phoenix} applications (\texttt{histogram} and \texttt{linear\_regression}). Among them, the first two applications actually have false sharing caused by the allocator. Five applications with conflict misses are collected from CCProf~\cite{CCProf}: \texttt{ADI}~\cite{adi}, \texttt{HimenoBMT}~\cite{himenobmt}, \texttt{Kripke}~\cite{Kripke}, \texttt{MKL-FFT}~\cite{mkl}, and \texttt{NW}~\cite{nw}. \texttt{TinyDNN}~\cite{tinydnn} is not included, since we did not observe conflict misses and the change (based on CCProf~\cite{CCProf}) did not improve the performance. We also include \texttt{irs}~\cite{irs} and \texttt{SRAD}~\cite{SRAD} applications that were employed by ArrayTool~\cite{ArrayTool} to evaluate capacity misses. 
Note: to reproduce false sharing on our machine, \texttt{histogram} processes a special BMP file adapted from the original one that all of the red values are set to 0 and the blue values are set to 255.  For \texttt{linear\_regression}, we also use the \texttt{volatile} keyword for the \texttt{args} variable in order to avoid the optimization of the compiler. For \texttt{HimenoBMT}, the grid size is medium and the number of integration is 80. \texttt{NW}'s matrix dimension is set to be $16384\times16384$, and its penalty is set to be 10. 

%For \texttt{TinyDNN}, we evaluted the CIFAR model by CIFAR10 dataset. Other applications are performed with their default inputs.

\paragraph{Evaluated Allocators: } To evaluate \CP{}'s detection on issues introduced by allocators, we evaluate on two widely-used allocators, \texttt{Glibc-2.28} and  \texttt{TCMalloc-4.5.3}. \texttt{Glibc-2.28} includes the default allocator in our machine, and \texttt{TCMalloc} is a widely-used allocated designed by Google~\cite{tcmalloc}.

\paragraph{Comparison:} We compare \CP{} with two state-of-art tools in effectiveness, performance, and memory consumption. One is CCProf which detects cache conflict misses~\cite{CCProf}, and the other one is Feather for false sharing detection~\cite{Feather}.  We have difficulty running ArrayTool~\cite{ArrayTool} successfully, which is the reason why ArrayTool is not included for comparison. For these tools, we use their default sampling rates used for their evaluation. 
%: \TP{1212 for CCProf (except) and 1 million for Feather.}

\subsection{Effectiveness}
\label{sec:effectiveness}

We list the effectiveness results of \CP{}'s detection in Table~\ref{tbl:effectiveness}. Overall, \textit{\CP{} reports all known bugs and detects 9 new bugs, while fixing the reported bugs achieves the performance improvement between 3\% and 3788\%}. Some applications with capacity misses cannot be easily fixed, marked as ``?'' in the ``Improve'' column. This also concurs with our discussion in Section~\ref{sec:misstype} that not all capacity misses can be fixed easily. \CP{} correctly identifies all types of bugs, except \texttt{bug 7} in \texttt{ADI}. The type is identified by \CP{} as capacity miss, but it is actually conflict miss. Based on our investigation, the failure of the identification is caused by the skids of the PMU hardware~\cite{skids}. The PMU hardware fails to pinpoint the exact instruction with the sampled cache miss, with the distance of one instruction. Therefore, \CP{} actually captures the access pattern of an instruction different from the one with cache misses, which does not have the pattern of conflict misses. However, our observation that \textit{``an instruction's access pattern is not changed during the whole execution''} still holds. 

Note that although \texttt{streamcluster} has been reported by previous tools with a false sharing issue, but achieving no performance improvement after fixing the bug as suggested by previous tools~\cite{Sheriff}. \CP{} successfully avoids the report of this bug, therefore, preventing programmers to spend the effort on this bug. In contrast, Feather still reports this insignificant bug, which is the reason why it is marked as ``\checkmark\kern-1.1ex\raisebox{.7ex}{\rotatebox[origin=c]{125}{--}}''. Feather cannot report the origin of false sharing in both \texttt{cache-scratch} and \texttt{cache-thrash}, which are allocator-induced conflict misses. Similarly, although CCProf reports conflict misses of \texttt{raytrace}, but it fails to identify as an allocator-induced miss. 
%Overall, \CP{} reports \textbf{three new} bugs that are unable to be accurately reported by existing tools. We further show some case studies of these bugs as follows. 
%, and one application conflict miss (\texttt{SRAD}). 
%Further, it skips one minor false sharing without significant performance impact (\texttt{streamcluster}). 
\begin{table*}
\footnotesize
\centering {
%\scalebox{0.8}{
\begin{tabular}{cclr|ccc|c}
\hline
Category   &  Index &   Application & Improve & CCProf & Feather & \textbf{CachePerf} & New \\
\hline
\multirow{5}{*}{False Sharing} & 1 & cache-scratch*              & 1007\%  & \xmark      & \checkmark\kern-1.1ex\raisebox{.7ex}{\rotatebox[origin=c]{125}{--}}    & \checkmark        & \checkmark  \\
&2 & cache-thrash*                     & 3788\%  & \xmark      & \checkmark\kern-1.1ex\raisebox{.7ex}{\rotatebox[origin=c]{125}{--}}    & \checkmark        & \checkmark  \\
& 3 & histogram           & 117\%   & \xmark      & \checkmark      & \checkmark        &     \\
& 4 & linear\_regression                  & 712\% & \xmark & N/A   & \checkmark&  \\
& 5 & streamcluster       & 0\%   & \xmark & \checkmark\kern-1.1ex\raisebox{.7ex}{\rotatebox[origin=c]{125}{--}} & \checkmark&  \\
\hline
\multirow{9}{*}{Conflict Miss} & 6 & ADI     & 246\%   & \checkmark     & \xmark       & \checkmark        &     \\
& 7 & ADI       & 18\%    & \xmark      & \xmark       &  \checkmark\kern-1.1ex\raisebox{.7ex}{\rotatebox[origin=c]{125}{--}}      &  \checkmark   \\
& 8 &  HimenoBMT          & 964\%   & \checkmark     & \xmark      & \checkmark        &     \\
& 9 & Kripke                & 7\%     & N/A     & N/A      & \checkmark        &     \\
& 10 & MKL\_FFT            & 52\%  & \checkmark& \xmark   & \checkmark&  \\
& 11 & NW             & 245\%   & \checkmark     & \xmark      & \checkmark        &     \\
& 12 & raytrace*      & 27\%    & \checkmark\kern-1.1ex\raisebox{.7ex}{\rotatebox[origin=c]{125}{--}}   & \xmark      & \checkmark        & \checkmark  \\
& 13 & SRAD              & 748\%   & \checkmark     & \xmark      & \checkmark        &   \checkmark  \\
& 14 & swaptions       & 3\%   & \xmark& \xmark   & \checkmark& \checkmark \\
\hline
\multirow{5}{*}{Capacity Miss} & 15 & bodytrack         & ?       & \xmark     & \xmark      & \checkmark        & \checkmark    \\
& 16 & canneal         & ?       & \xmark     & \xmark      & \checkmark        &  \checkmark   \\
& 17 & IRS           & 33\% & \xmark     & \xmark      & \checkmark        &     \\
& 18 & SRAD         & 12\% & \xmark     & \xmark      & \checkmark        &     \\
& 19 & streamcluster    & ?       & \xmark     & \xmark      & \checkmark        &  \checkmark  \\
\hline
\end{tabular}
}
%}
\captionsetup{justification=justified}
\caption{This table lists  applications with cache misses. For applications marked with *,  \texttt{cache-scratch}, \texttt{cache-thrash} have allocator-induced false sharing, and \texttt{raytrace} has allocator-induced conflict misses.  Column ``Improve'' lists the performance improvement after fixes based on information provided by \CP{}, where column ``New'' indicates whether it is first discovered by \CP{}. Further,  ``\checkmark'' indicates the tool correctly detects the issue, ``\checkmark\kern-1.1ex\raisebox{.7ex}{\rotatebox[origin=c]{125}{--}}'' indicates an imperfect report,  ``\xmark'' indicates a failed detection, and ``N/A'' indicates that the corresponding application crashes or deadlocks when running with the tool. Note that applications marked as ``?'' in ``Improve'' cannot be fixed easily, which confirms our discovery in Section~\ref{sec:misstype}.  \label{tbl:effectiveness}}
\vspace{-0.1in}
\end{table*}

%\TP{could we find more issues using the existing tools?}
\subsubsection{Conflict Misses of Applications}
\label{sec:applicaionConflict}
\CP{} could correctly report all known conflict misses, including \texttt{ADI}, \texttt{HimenoBMT}, \texttt{Kripke}, \texttt{MKL-FFT}, and \texttt{NW}.  These bugs can be fixed by switching the order of loops (\texttt{Kripke}) and using the padding (others). 

\begin{figure*}
\centering
\includegraphics[width=5.4in]{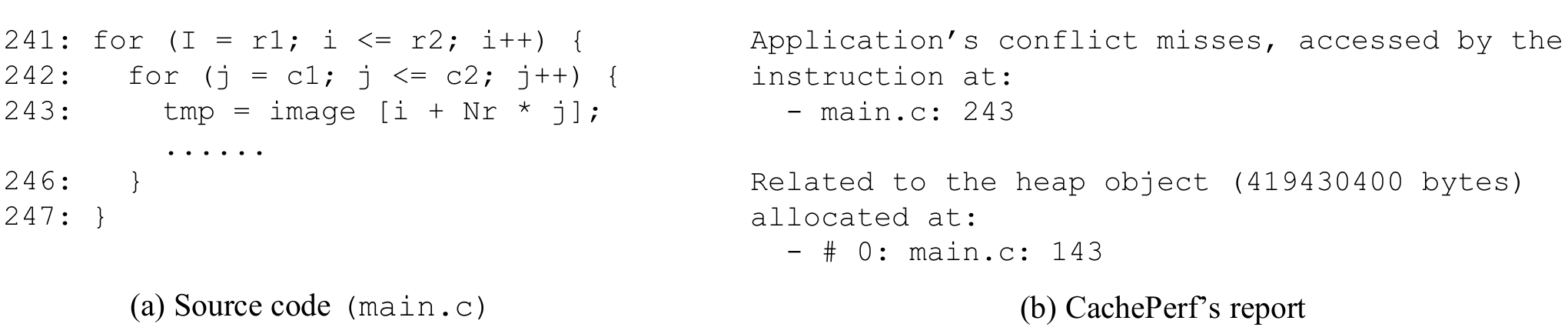}
\caption{The conflict miss in \texttt{SRAD}, which can be fixed easily by switching the loops of line 241 and 242.
\label{code:sradconflict}}
\end{figure*}

\CP{} further detects three unknown conflict misses, in \texttt{ADI}, \texttt{SRAD}, and \texttt{swaptions}. The bug of the \texttt{SRAD} application is shown in Fig.~\ref{code:sradconflict}, which can be detected by CCProf. \CP{} reports that line 243 of \texttt{main.c} introduces around 64\% of load misses. As shown in Fig.~\ref{code:sradconflict}(a), \texttt{SRAD} uses two nested for loops to calculate the sum for every pixel in the image ROI.  By simply switching these two loops, we improve the performance by 748\%. 
%In line 243, SRAD uses i + Nr*j to index the image array. However, with current loop layout j changes faster than i, making the image array access noncontinuous. This means subsequent memory access cannot effectively use the previously loaded cache line and will likely cause conflict with other cache lines. 

\subsubsection{Allocator-Induced Conflict Misses} 

\begin{figure*}[!ht]
\centering
\includegraphics[width=5.4in]{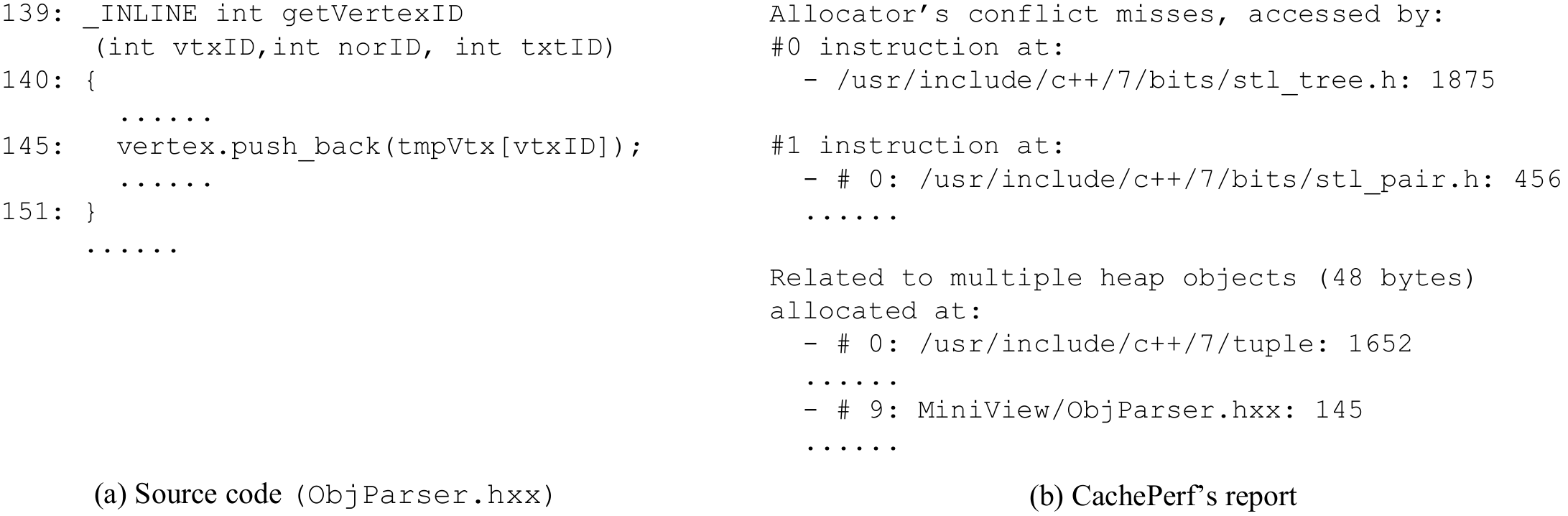}
\caption{\CP{} reports an allocator-caused conflict miss in \texttt{raytrace}
\label{code:raytrace}}
\end{figure*}

\CP{} also detects a serious conflict miss in \texttt{raytrace} caused by the default allocator--\texttt{glibc-2.28}, as shown in Fig.~\ref{code:raytrace}. The report can be seen in Fig.~\ref{code:raytrace}(b). For this problem, the default \texttt{glibc-2.28} happens to allocate many 48-byte objects mapping to the same cache set, causing conflict misses. \texttt{TCMalloc} does not have this issue, which runs about 27\% faster on this application than \texttt{glibc-2.28}. We further confirm whether there exists a systematic method in \texttt{TCMalloc} to prevent such an issue. We find that \texttt{TCMalloc} always requests two pages at a time, then allocates objects (48 bytes) continuously, and skips non-used bytes in the end. This mechanism luckily avoids conflict misses of \texttt{raytrace} application. In summary, allocator-induced conflict misses are not easy to prevent from the design of the allocator. This also shows the importance of \CP{} that could help identify the root cause of performance slowdown. After finding out the issue, programmers may switch to a different allocator, or change the application by introducing unnecessary allocations inside or changing the alignment of the related structure.  

\subsubsection{Capacity Misses of Applications}
We borrowed some buggy applications from  ArrayTool~\cite{ArrayTool}, including \texttt{IRS}~\cite{irs} and \texttt{SRAD}~\cite{SRAD}. The paper also reports serious issues in a specific version of \texttt{LULESH}~\cite{karlin2012lulesh}. However, we cannot find the exact source code, which is the reason why \texttt{LULESH} is not included. Besides these applications, \CP{} also detects unknown capacity misses in \texttt{bodytrack}, \texttt{canneal}, and \texttt{streamcluster}, which has been confirmed by us manually. However, as mentioned in Section~\ref{sec:misstype}, not all capacity misses could be fixed easily. 

\begin{figure*}[!h]
\centering
\includegraphics[width=5.4in]{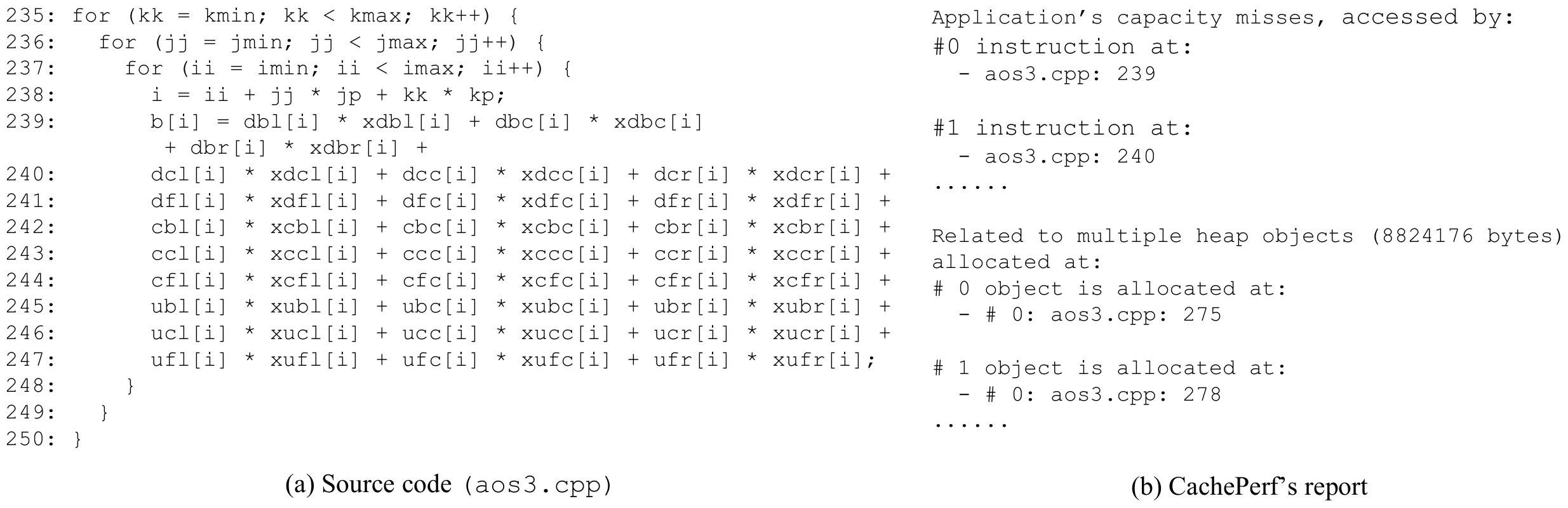}
\caption{Reported capacity miss in \texttt{IRS}
\label{code:irs}}
\end{figure*}

As shown in Table~\ref{tbl:effectiveness}, \CP{} successfully reports capacity misses hidden in both \texttt{IRS} and \texttt{SRAD}. As an example, the \texttt{IRS}'s source code and report are shown in Fig.~\ref{code:irs}.  \texttt{IRS}'s capacity misses occur in line 239 of \texttt{aos3.cpp}, although \texttt{addr2line} actually reports lines between 239 and 247. This statement accesses many objects of the same size (88824176 bytes), e.g., dbl, xdbl, dbc. Since every object has exactly the same access pattern, these accesses should be grouped together. Using the suggested fix strategy~\cite{ArrayTool}, the performance can be improved by 32.7\%.

% \begin{minipage}[c]{0.95\linewidth}
% \begin{lstlisting}[basicstyle=\small, frame=single, caption={\CP{} reports capacity misses in \texttt{IRS}}, captionpos=b, label={code:irs}] 
% 234: #pragma omp parallel for private(jj,ii,i)
% 235: for (kk = kmin; kk < kmax; kk++) {
% 236:   for (jj = jmin; jj < jmax; jj++) {
% 237:     for (ii = imin; ii < imax; ii++) {
% 238:       i = ii + jj * jp + kk * kp;
% 239:       b[i] = dbl[i] * xdbl[i]  + dbc[i] * xdbc[i] 
%                 + dbr[i] * xdbr[i] +
% 240:       dcl[i] * xdcl[i] + dcc[i] * xdcc[i] + dcr[i] * xdcr[i] +
% 241:       dfl[i] * xdfl[i] + dfc[i] * xdfc[i] + dfr[i] * xdfr[i] +
% 242:       cbl[i] * xcbl[i] + cbc[i] * xcbc[i] + cbr[i] * xcbr[i] +
% 243:       ccl[i] * xccl[i] + ccc[i] * xccc[i] + ccr[i] * xccr[i] +
% 244:       cfl[i] * xcfl[i] + cfc[i] * xcfc[i] + cfr[i] * xcfr[i] +
% 245:       ubl[i] * xubl[i] + ubc[i] * xubc[i] + ubr[i] * xubr[i] +
% 246:       ucl[i] * xucl[i] + ucc[i] * xucc[i] + ucr[i] * xucr[i] +
% 247:       ufl[i] * xufl[i] + ufc[i] * xufc[i] + ufr[i] * xufr[i];
% 248      }
% 249    }
% 250 }

% \end{lstlisting}
% \end{minipage}

% \begin{minipage}[c]{0.95\linewidth}
% \begin{lstlisting}[basicstyle=\ttfamily, frame=single, caption={\CP{} report for capacity misses in \texttt{irs}}, captionpos=b, label={fig:irs}]
% Application's capacity Misses by 
% susceptible instructions:
%     - aos3.cpp: 239-247

% The related heap object (with size 8824176 bytes) is allocated at:
%     - #0: aos3.cpp:276-304

% \end{lstlisting}
% \end{minipage}

Note that \CP{} cannot report \texttt{SRAD}'s capacity miss in the original version when the conflict miss (as shown in Fig.~\ref{code:sradconflict}) is the dominant performance issue. We also confirmed that applying the suggested fix by ArrayTool~\cite{ArrayTool} achieves almost no performance improvement. In fact, this actually illustrates the effectiveness of \CP{} as its rule-based filtering mechanism  avoids reporting minor issues. After fixing the conflicting miss of \texttt{SRAD}, then \CP{} could successfully report the capacity miss. After fixing the report bug, \texttt{SRAD}'s performance is improved by 12.4\% finally. 

\subsubsection{Coherency Misses (FS) of Applications}
\begin{figure*}[h!]
\centering
\includegraphics[width=5.4in]{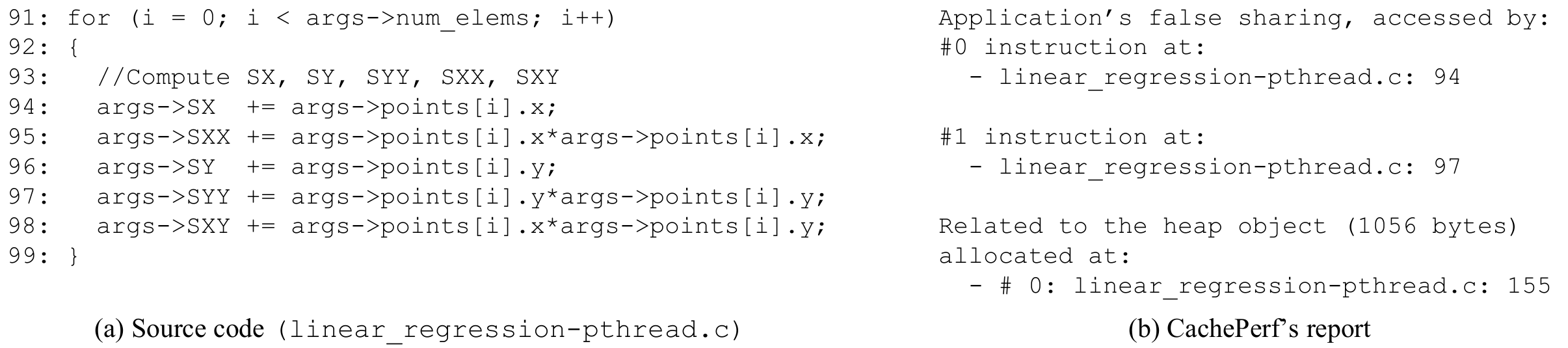}
\caption{\CP{} reports false sharing in \texttt{linear\_regression}
\label{code:regression}}
\end{figure*}

For coherency misses of applications, we utilize three known buggy applications to evaluate \CP{}'s effectiveness, including \texttt{histogram}, \texttt{linear\_re-}
\texttt{gression}, and \texttt{streamcluster}. 
\CP{} successfully detects these issues latent in  \texttt{histogram} and \texttt{linear\_regression}, similar to existing work~\cite{Sheriff, Feather}. We show the source code and \CP{}'s report of \texttt{linear\_regression} in Fig.~\ref{code:regression}. This is a known bug that the structure of \texttt{args} is not aligned to 64 bytes (but only 52 bytes instead). As a result, thread 1 will access the same cache line as thread 2. By simply aligning the related structure, the performance can be improved by 712\%. Different from existing tools, \CP{} will not report the issue of \texttt{streamcluster}, although it was reported to have false sharing for the \texttt{work\_mem} object~\cite{Sheriff}. Based on existing work, we fixed the false sharing by using the padding and observed the reduction of cache misses. However, we do not observable performance impact with this change, less than 1\%. That is, \CP{} successfully excludes the insignificant issue, avoiding the waste of manual effort. In contrast, Feather still reports this false sharing of \texttt{streamcluster}, although it only imposes little performance impact. 

\vspace{-1mm}

\subsubsection{Allocator-Induced False Sharing}

\begin{figure*}[h!]
\centering
\includegraphics[width=5.4in]{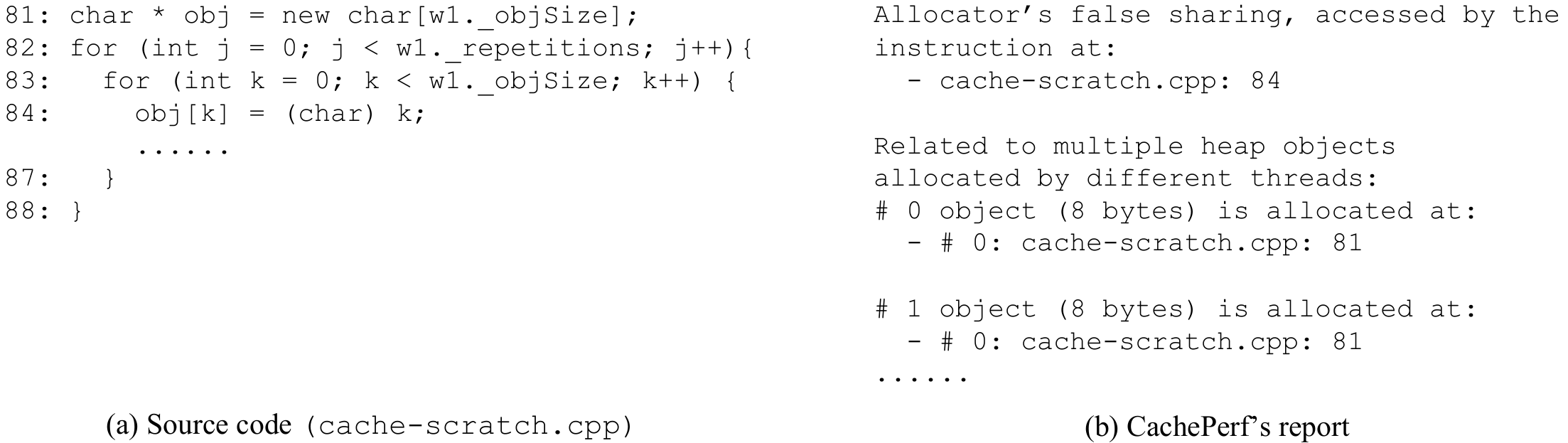}
\caption{\CP{} reports coherence misses in \texttt{cache-scratch}
\label{code:cache-scratch}}
\end{figure*}

When using the default allocator, \CP{} also reports allocator-induced false sharing as shown in Fig.~\ref{code:cache-scratch}. \CP{} infers allocator-induced false sharing as more than two objects allocated by different threads are located in the same cache line, as shown in Fig.~\ref{code:cache-scratch} (b). A simple solution is to change the alignment of the structure related to \texttt{obj}, which improves the performance by 1007\%. \CP{} also reports some serious allocator-caused false sharing issues for both \texttt{cache-scratch} and \texttt{cache-thrash} with the \texttt{TCMalloc} allocator.  There will be 3788\% performance improvement using the padding. Although allocator-caused false sharing is a bug of the allocator design, it can be prevented by changing the application itself. In addition, users could switch to a new allocator to fix such issues. \CP{} provides helpful information that could help fix such bugs. 

% \begin{figure}[!h]
% \centering
% \includegraphics[width=3in]{paper/Figure/report2.pdf}
% \captionsetup{justification=justified}
% \caption{\CP{} report for coherence misses in \texttt{cache-scratch}.
% \label{fig:cache-scratch}}
% \end{figure}
%\begin{figure}

%\vspace{-1mm}

\textbf{Comparing with Other Tools:}
% \subsubsection{Applications with Conflict Misses}
Overall, \CP{} shows \textbf{three obvious advantages} when compared with existing tools. First, \CP{} can detect multiple types of cache misses, while others could only report a specific type of cache misses. Note that the other tools are mutually exclusive, forcing programmers to use them one after the other. Second, \CP{} is the only tool that identifies the performance issues introduced by the memory allocator, preventing programmers from wasting the unnecessary effort of improving applications but achieving no performance improvement. Finally, \CP{} is the only tool excluding minor issues with little performance impact, saving users' time.

\subsection{Performance Overhead}
\label{sec:perf}

\begin{figure*}[h!]
\centering
\includegraphics[width=5.4in]{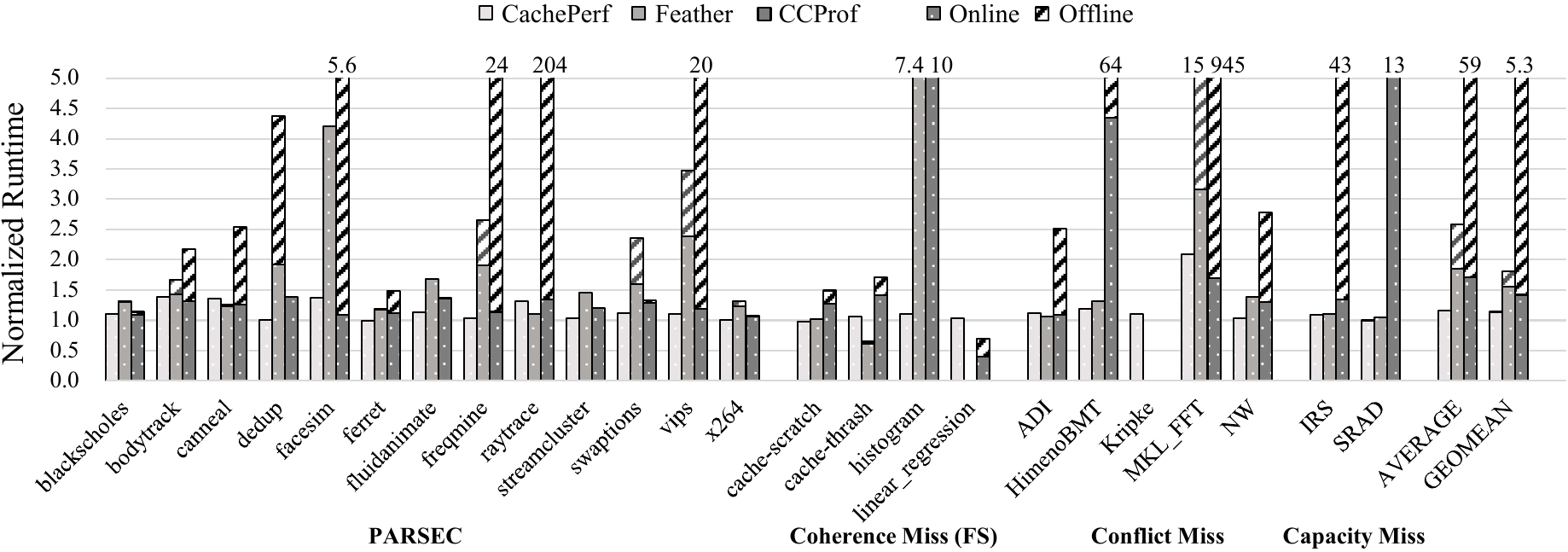}
% \captionsetup{justification=justified}
\caption{The performance of \CP{}, CCProf, and Feather, \\where the results are normalized to the default setting (without running any tool). \label{fig:performance}}
\end{figure*}

We evaluated the performance overhead of \CP{}, CCProf, and Feather. Since CCProf and Feather have online and offline stages, we add their overhead of two stages together. The results (with the AVERAGE and GEOMEAN) are shown in Fig.~\ref{fig:performance}.

%\todo{Add description or even application characteristics in a table, which helps explain why \CP{} is slower on some applications. } 
On average (GEOMEAN), \CP{} introduces 14\% performance overhead, while CCProf's overhead is  $5.3\times$ and Feather's overhead is 80\% if considering both online profiling and offline analysis. 
Even only considering their online profiling, \CP{} is still faster than both CCProf and Feather. Based on our understanding, \CP{}'s ratio-based mechanism helps reduce much unnecessary overhead by pruning sporadic cache misses, but without compromising its effectiveness.  \CP{}'s specific data structures also help reduce the overhead. 
%Based on our observation, \CP{}'s Miss Ratio Checker saves much runtime by not processing discrete information of issues with little performance impact. 
\CP{}'s hybrid sampling technique that combines with both coarse-grained and fine-grained sampling also balances the accuracy and the overhead.  On one hand, its coarse-grained sampling works as a filter that allows it to focus on only a few susceptible instructions, avoiding installing unnecessary breakpoints. On the other hand, the breakpoints effectively ensure the precision of the tool even with a low PMU-based sampling rate.

\texttt{MKL\_FFT} is the only application with an overhead higher than 100\%. We confirmed that more than 80\% of its overhead is spent in its reporting phase, which could be placed offline if necessary. This application has involved a big amount of cache sets, heap objects, and instructions. For instance, \CP{} requires to invoke the expensive \texttt{addr2line} to obtain the line numbers for many lines. We are planning to reduce such overhead with  heuristics in the future. 

\CP{} introduces 36\% performance overhead for \texttt{canneal}. The basic reason is that \texttt{canneal} has a great number of allocations (about 1.3 million per second), where keeping the information of these objects adds significant overhead (and memory overhead). Similarly, \CP{} introduces high overhead for keeping and updating the information of objects for \texttt{raytrace}, as there are around 500 thousand memory allocations each second. \CP{} introduces high overhead for \texttt{bodytrack} and \texttt{facesim} for a similar reason. 

%\CP{} also introduces 38\% performance overhead for \texttt{bodytrack}. \CP{} can introduce higher overhead by keeping more instructions and cache lines in its hash tables. 

%We also notice that \CP{} introduces 37\% performance overhead for \texttt{facesim}. However, we does not find irregular numbers from that statistic. We will further investigate that in the future work.
%However, overall, \CP{}'s overhead is still acceptable, considering its uniqueness and powerful functionalities. 

We further checked the reason why Feather runs faster with \texttt{cache-thrash} and CCProf runs faster with \texttt{linear\_regression}. Based on our investigation, Feather allocates some memory from the default memory allocator for its internal usage, which happens to alleviate the false sharing issue introduced by the allocator. Similarly, CCProf's memory usage also changes the starting address of the false sharing object, reducing the severity of false sharing. That is, they should impose higher overhead if such lucky cases are excluded. We also observe that CCProf's offline phase is very expensive, e.g., \texttt{MKL\_FFT}, which could be as much as $945\times$ higher. In contrast, \CP{} does not have the hidden overhead for the offline analysis, which could report cache misses immediately after the execution or when receiving the signal from users (good for long-running applications).

%Compared with \texttt{histogram} that also has short runtime, \CP{} catches more instructions with cache misses in \texttt{ADI} and tries to convert them into source codes via external \texttt{addr2line} commands. \CP{} also prints metadata of more cache sets and objects with cache misses in \texttt{ADI} than in \texttt{histogram}, which cause the reporting stage of \texttt{ADI} more costly.
\subsection{Memory Overhead}
\label{sec:memory}

We also evaluated the memory overhead of \CP{}, CCProf, and Feather, as shown in Table~\ref{tbl:memory}. Since CCProf crashed for \texttt{Kripke}, and Feather encountered the deadlock for \texttt{Kripke} and \texttt{linear\_regression}, these applications are marked ``N/A'' in the table. 

\begin{table}
\centering
\footnotesize
% \scalebox{0.92}{
\setlength{\tabcolsep}{0.4em}
\begin{tabular}{llrrrr}
\hline
Category        & Application &
 Default &
  \textbf{\CP{}} &
  CCProf & 
Feather \\ 
                               \hline 
\multirow{13}{*}{PARSEC}         & blackscholes            & 614  & 632  & 1850 & 622  \\ 
                                 & bodytrack               & 34   & 70   & 1397 & 44    \\ 
                                 & canneal                 & 851  & 1728 & 2089 & 860   \\ 
                                 & dedup                   & 1513 & 1647 & 2811  & 1549  \\ 
                                 & facesim                 & 311  & 390  & 2776  & 330    \\ 
                                 & ferret                  & 108  & 142  & 1343 & 116   \\ 
                                 & fluidanimate            & 209  & 234  & 2675 & 218   \\ 
                                 & freqmine                & 1280 & 1312 & 3736 & 1289  \\ 
                                 & raytrace                & 1287 & 1319 & 3213 & 1295  \\ 
                                 & streamcluster           & 112  & 238  & 2572  & 122  \\ 
                                 & swaptions               & 7    & 36   & 1244 & 16    \\ 
                                 & vips                    & 55   & 80   & 1234 & 108   \\ 
                                 & x264                    & 482  & 514  & 1711 & 510 \\ \hline 
\multirow{4}{*}{\begin{tabular}[c]{@{}c@{}}Coherence Miss\\ (False Sharing)\end{tabular}} &
  cache-scratch &
  3 &
  22 &
  1551 &
  11  \\  
                                 & cache-thrash            & 4    & 28   & 1810 & 11  \\ 
                                 & histogram          & 1344 & 1362 & 2574 & 1336 \\ 
                                 & linear\_regression & 1956 & 1974 & 3185 & N/A  \\ \hline 
\multirow{5}{*}{Conflict Miss} & ADI                     & 514  & 528  & 1743 & 520  \\ 
                                 & HimenoBMT               & 225  & 302  & 1455 & 232\\ 
                                 & Kripke                  & 301  & 360  & N/A   & N/A  \\ 
                                 & MKL-FFT                & 261  & 283  & 1490 & 268   \\ 
                                 & NW                      & 2050 & 2068 & 3279 & 2058 \\ 
                                 \hline 
       \multirow{2}{*}{Capacity Miss} 
                                 & IRS               & 248 & 342  & 3118 & 256\\ & SRAD                     & 2404  & 2420  & 4869 & 2412  \\   \hline      \textbf{TOTAL} &                     &  16174   &   18032  & 53726   &  14182   \\                    
%\multicolumn{2}{c}{\textbf{AVERAGE}}                     &     &    190\%    &    5630\%     &    133\%      \\ 
\textbf{GEOMEAN}         &        &         &    151\% &    889\%     &  122\%   \\ \hline
\end{tabular}
% }
% \captionsetup{justification=justified}
\caption{The memory consumption (MB) of \CP{},  CCProf ,  and Feather. \\Column ``Default' lists  memory footprints of applications when running alone. \label{tbl:memory}}
\end{table}

In total, \CP{} adds around 11\% memory consumption, although its average overhead is around 51\%. When only considering the online stages of CCProf and Feather, \CP{}'s memory overhead is significantly better than CCProf, but slightly worse than Feather. However, if the offline stage is also considered when using the maximum memory consumption of both stages, then \CP{} has the smallest memory consumption. %Similar to the performance overhead, \CP{}'s Miss Ratio Checker and the hybrid sampling save memory from saving unimportant data, and help reduce the memory overhead. Therefore, \CP{} keeps reasonable memory overhead on average, though it has provided more functions and helper information than other tools.

Table~\ref{tbl:memory} shows that \CP{} introduces high memory overhead for applications with small memory footprints, such as \texttt{swaptions}, \texttt{cache-scratch}, and \texttt{cache-thrash}. Based on our observation, the overhead is introduced by \CP{}'s initialization overhead for its pre-defined hash tables. However, \CP{} only introduces around 19\% memory overhead on average for applications with large footprints (e.g., $>100MB$). Considering the functionalities provided by \CP{}, we believe that the memory overhead of \CP{} is reasonable and acceptable.

\subsection{Impact of Configurations}
\label{sec:thresholdImpact}

In this section, we investigate the performance and effectiveness impacts of different configurations using all PARSEC applications. We investigate the impact of the sampling rate, thresholds of Miss Ratio Checker, breakpoint configurations, and access/miss ratio. 

\subsubsection{Sampling Rate}

We evaluated three sets of sampling periods as shown in Table~\ref{tbl:periods}. With its default setting (marked as bold), \CP{}'s GEOMEAN performance and memory overhead are 14\% and 48\% separately. When the sampling frequencies are 10 times lower than the default setting, the performance overhead is 10\% and the memory overhead is 47\%. However, as shown in the ``CP1'' column of Table~\ref{tbl:samplingEffect}, \CP{} will miss 10 out of 19 issues. When the sampling frequencies are 10 times higher than the default setting, the performance and memory overhead are increased to 18\% and 79\% correspondingly, but do not report more issues.
%Both overhead increases since \CP{} handles about 10 times more sampled events.
Overall, \CP{}'s default sampling periods keep a good balance between performance and effectiveness.
%Further, we confirmed that the effectiveness does not improve with more frequent sampling. 

\begin{table}[!h]
\scalebox{0.85}{
\begin{tabular}{llllll}
\hline
 &
  \begin{tabular}[c]{@{}l@{}}L:200K, S: 500K \\ Miss Ratio: 0.5\%\end{tabular} &
  \begin{tabular}[c]{@{}l@{}}L:20K, S: 50K \\ Miss Ratio: 2.5\%\end{tabular} &
  \textbf{\begin{tabular}[c]{@{}l@{}}L:20K, S: 50K\\ Miss Ratio: 0.5\%\end{tabular}} &
  \begin{tabular}[c]{@{}l@{}}L:2K, S: 5K \\ Miss Ratio: 0.5\%\end{tabular} &
  \begin{tabular}[c]{@{}l@{}}L:20K, S: 50K \\ Miss Ratio: 0\%\end{tabular} \\ \hline
Performance &
  10\% &
  12\% &
  \textbf{14\%} &
  18\% &
  18\% \\
Memory &
  47\% &
  45\% &
  \textbf{48\%} &
  79\% &
  70\% \\ \hline
\end{tabular}
}
\captionsetup{justification=justified}
\caption{This table lists the performance and memory overhead under different sampling periods (``L'' is the load sampling period and ``S'' is the store sampling period) and different thresholds of the Miss Ratio Checker (``Miss Ratio'') using all PARSEC applications. The middle column (in bold) is the default configuration. \label{tbl:periods}}
\end{table}

\setlength\tabcolsep{1.2pt}
\begin{table}
\centering {
\scalebox{0.85}{
\begin{tabular}{cl|ll|ccc}
\hline
Index & Application    & Instructions                      & Allocation Callsite        & CP & CP1 & CP2 \\ \hline
1     & cache-scratch* & cache-scratch.cpp: 84       & cache-scratch.cpp: 81      & \checkmark  & \checkmark   & \checkmark   \\
2     & cache-thrash*  & cache-thrash.cpp: 84        & cache-thrash.cpp: 75       & \checkmark  & \checkmark   & \checkmark   \\
3     & histogram      & histogram-pthread: 126, 132 & histogram-pthread: 231     &\checkmark  & \xmark   & \checkmark\\
4  & linear\_regression & linear\_regression-pthread.c: 94, 97      & linear\_regression-pthread.c: 155           & \checkmark  & \xmark   & \checkmark \\
5  & streamcluster      & streamcluster.cpp: 1005, 1015, 1098, 1099 & streamcluster.cpp: 985                      & \checkmark  & \checkmark   & \checkmark \\ \hline
6     & ADI            & adi.c: 104                  & utilities/polybench.c: 524 & \checkmark  & \checkmark   & \checkmark   \\
7     & ADI            & adi.c: 109                  & utilities/polybench.c: 524 & \checkmark\kern-1.1ex\raisebox{.7ex}{\rotatebox[origin=c]{125}{--}}  & \checkmark\kern-1.1ex\raisebox{.7ex}{\rotatebox[origin=c]{125}{--}}   & \checkmark\kern-1.1ex\raisebox{.7ex}{\rotatebox[origin=c]{125}{--}}   \\
8     & HimenoBMT      & himenoBMTxpa.c: 295-316     & himenoBMTxpa.c: 231        & \checkmark  & \xmark   & \checkmark   \\
9     & Kripke         & Grid.cpp: 262               & SubTVec.h: 54              & \checkmark  & \xmark   & \xmark   \\
10 & MKL\_FFT           & MKL Library                               & basic\_dp\_xx\_2d\_4096.c: 88, 95 & \checkmark  & \xmark   & \checkmark \\ 
%basic\_dp\_complex\_dft\_2d\_4096.c
11    & NW             & needle.cpp: 130, 191, 290   & needle.cpp: 262, 263       & \checkmark  & \xmark   & \xmark   \\
12    & raytrace*      & C++ STL Library             & MiniView/rtview.cxx: 410   & \checkmark  & \checkmark   & \checkmark   \\
13    & SRAD           & main.c: 243                 & main.c: 143                & \checkmark  & \checkmark   & \checkmark   \\
14 & swaptions          & HJM\_SimPath\_Forward\_Blocking.cpp: 121  & nr\_routines.cpp: 168                       & \checkmark  & \xmark   & \xmark \\ \hline
15    & bodytrack      & ImageMeasurements.cpp: 43   & AsyncIO.cpp: 55            & \checkmark  & \xmark   & \checkmark   \\
16    & canneal        & C++ STL Library             & netlist.cpp: 236           & \checkmark  & \checkmark   & \checkmark   \\
17    & IRS            & aos3.cpp: 239-247           & aos.cpp: 275-304           & \checkmark  & \checkmark   & \checkmark   \\
18    & SRAD           & main.c: 312                 & main.c: 188-191            & \checkmark  & \xmark   & \xmark   \\
19    & streamcluster  & streamcluster.cpp: 652      & streamcluster.cpp: 1862    & \checkmark  & \xmark & \xmark  \\ \hline
\end{tabular}
}}
\captionsetup{justification=justified}
\caption{This table lists the effectiveness of \CP{} under different configurations. ``CP'' is the default setting, with the load and store sampling periods to be 20K and 50K separately. ``CP1'' has 10 times lower sampling frequencies ( 200K and 500K separately), but will miss 10 cases. In ``CP2'', its miss ratio checker uses a higher threshold (2.5\%), which will miss 5 cases. ``\checkmark'', ``\checkmark\kern-1.1ex\raisebox{.7ex}{\rotatebox[origin=c]{125}{--}}'', and ``\xmark'' have the same meaning as Table~\ref{tbl:effectiveness}.
%For applications marked with *,  \texttt{cache-scratch}, \texttt{cache-thrash} have allocator-induced false sharing, and \texttt{raytrace} has allocator-induced conflict misses. ``\checkmark'' indicates the tool correctly detects the issue, ``\checkmark\kern-1.1ex\raisebox{.7ex}{\rotatebox[origin=c]{125}{--}}'' indicates an imperfect report,  ``\xmark'' indicates a failed detection.  
\label{tbl:samplingEffect}}

\end{table}

\subsubsection{Threshold of Miss Ratio Checker}
\label{sec:thresholds}

We further investigate the impacts of different thresholds of the Miss Ratio Checker. \CP{} will handle all cache misses inside the buffers, when the cache miss ratio is larger than the pre-defined threshold. 
As described in Section~\ref{sec:missRatioChecker}, the default threshold is 0.5\%. In the default setting, \CP{}'s performance and memory overhead are 14\% and 48\% separately. When the threshold is increased to 2.5\%, indicating \CP{} will only handle all cache misses when there are 25 misses out of 1000 accesses, the performance overhead is 12\% and the memory overhead is 45\%, as shown in Table~\ref{tbl:periods}. However, as shown in ``CP2'' in Table~\ref{tbl:samplingEffect}, \CP{} will miss 5 issues under this configuration. Another setting is 0\%, indicating \CP{} will handle all cache misses in the buffer, the performance and memory overhead is 18\% and 70\% correspondingly. However, this setting does not report more issues. Overall, the default threshold of Miss Ratio Checker has a good balance between overhead and effectiveness.

\subsubsection{Breakpoint Configuration}

We also evaluated the overhead and effectiveness impacts of different breakpoint configurations. As discussed in Section~\ref{sec:breakpointHandler}, \CP{} collects at most 64 accesses from one selected instruction, and identifies the bug as the conflict miss when more than 8 accesses are landing on the same cache set. That is, \CP{} will remove the breakpoint on this instruction if 8 continuous accesses are from the same set.  Besides this default setting, we also evaluated using 4 or 16 accesses as the condition for identifying the conflict miss. We also evaluated different expiration time for the breakpoint installing on an instruction, such as 10ms and 1000ms, where the breakpoint will be installed for a new instruction. However, we do not observe a significant difference in overhead or effectiveness for different configurations. 

\subsubsection{Thresholds of Miss Rates}
As described in Section~\ref{sec:classifier}, \CP{} will skip the report if the number of load misses is less than 3\% of all load accesses and the number of store misses is less than 1\% of all store accesses. The goal is to exclude minor issues. To evaluate the correctness of the two thresholds, we checked the load and store miss rates of all evaluated applications. Overall, for applications with reported issues, as listed in Table~\ref{tbl:effectiveness}, their load or store miss rates are higher than the default thresholds. For applications where we do not observe significant issues (not listed in Table~\ref{tbl:effectiveness}), the load and store miss rates are both lower than these thresholds. Therefore, the current thresholds of miss rates are helpful to filter out minor issues and highlight significant issues of cache misses.

\section{Discussion}
\label{sec:discussion}
This section discusses the compatibility, thresholds, and limitation of \CP{}. 

\subsection{Compatibility}
\CP{} can be easily adapted to different hardware environments, such as cache with different cache line sizes or associativity. Currently, the cache-related parameters (e.g. cache line size, cache associativity) are listed in a configuration file. If users would like to use \CP{} for hardware with a different setting, they only need to change this configuration file. %Further, \CP{} relies on the PMU hardware to sample memory accesses, which could be affected when such hardware events are not available. However, based on our understanding, both X86 and ARM processors should support similar events, although it may require some engineering work to set up these hardware events.

\subsection{Configurable Thresholds}
\CP{} introduces some thresholds to control the sampling and the reporting. Such thresholds are confirmed to balance the overhead and effectiveness on the evaluated machine. In a different environment, users may need to change these thresholds. The thresholds used by \CP{} can be easily changed via compilation flags or environmental settings. 
%Users can change these thresholds to balance the overhead and effectiveness of \CP{}. The current settings are based on the evaluated machine, 
%For example, to report an instruction, the default threshold of memory accesses is 0.01\% of the total accesses, and the threshold of cache misses is 1\% of the total misses; when users only want to identify the issues with the most significant slowdowns, they can increase the thresholds; when users want to identify more minor issues, they can decrease these two thresholds.

\subsection{Limitation}
\label{sec:limit}

 \CP{} utilizes the hardware-based sampling techniques to perform the profiling, which has the benefits that do not need to change the programs and imposes little performance overhead. However, the setting of the PMU-based sampling may require some slight changes on different machines with different implementations. Since the PMU-based sampling and the breakpoint-based sampling are generally supported by different hardware architecture, the proposed techniques should be applicable for different hardware. 
 
 %Further, \CP{} may not work for the cloud computing environemnt when 
\section{Related Work}

\label{sec:related}

We discuss the related work based on the type of cache misses in the following. Although some tools, such as perf~\cite{perf}, oprofile~\cite{oprofile}, different Pin tools~\cite{jaleel2008cmp, MultiCacheSim}, or cachegrind (one tool inside Valgrind~\cite{Valgrind}), could report the percentage of cache misses in the lines of code, they cannot identify the type of cache misses. Therefore, they are not the focus of this paper. In the following, we only list tools that could identify the type of cache misses. 
%Existing work aiming to identify cache-related performance issues can be classified into two categories. One focuses on cache conflict misses, while another focuses on performance issues caused by cache coherency traffic, including false and true sharing. This section discussed these two categories separately. 

\paragraph{Detecting Capacity Misses:} 
Tao et al. propose a cache simulator that can identify cache capacity misses using the reuse distance for each memory access~\cite{cacheSimulation}. Nikos et al. propose another cache simulation methodology~\cite{nikoleris_delorean_2018}. Both cache simulators could study cache behaviors under various cache configurations, but neither of them can be used as an online profiling tool due to their prohibitive overhead. Delorean~\cite{delorean} improves the simulation efficiency, and identifies cache capacity misses by the number of distinct memory accesses since the last access to the observed cache line. However, it is still a simulation technique that requires the inspection of every memory access, which is slow too. ArrayTool focuses on a special type of capacity misses caused by multiple arrays~\cite{ArrayTool}. It utilizes the PMU-sampling to collect memory samples and determines candidate arrays by the combination of array affinities and array's access patterns.  

\paragraph{Detecting Conflict Misses:}
Cache simulators detect conflict misses by simulating the cache behavior based on the memory trace~\cite{lebeck1994cache, zhang2001mathematical}, but they are too slow to be used for online profiling. 
%CCProf detects cache conflict misses based on the observation that ``a relatively larger portion of cache misses in a subgroup of the total cache sets over the others indicates conflicts in those cache sets''~\cite{CCProf}. However, as mentioned in Section~\ref{sec:intro}, this observation is not correct that could introduce both false positives and false negatives. Further, 
CCProf proposes to employ Re-Conflict Distance to filter out cache sets with low RCD~\cite{CCProf}, based on address sampling. However, CCProf may introduce high performance overhead due to the use of a low sampling rate to capture RCD. As shown in Fig.~\ref{fig:performance}, the overhead of CCProf can be as much as 945 times. In contrast, \CP{} imposes significantly less overhead while could identify different types of cache misses.   
%Also, CCProf does not propose a good metric to evaluate the seriousness of performance issues due to the following two reasons. First, the percentage of cache misses on a specific cache line cannot justify the seriousness of cache-conflict. 
%Second, CCProf utilizes two obscure metrics that should be combined together to evaluate the seriousness. As shown in the paper, if the re-conflict distance is low and cache-miss contribution is high, then it may have a performance impact. 
%Instead, \CP{} proposes a cache-conflict score to evaluate the seriousness directly, which integrates cache access ratio, cache misses ratio, and re-conflict distance together. Given this metric, there is no need to rely on manual expertise and avoid false positives. 
\paragraph{Detecting Cache Coherency Misses:}
There exist multiple types of tools that could detect cache coherence issues, mostly focusing on false sharing. Some tools are relying on binary instrumentation~\cite{Zhaoqin}, compiler-based instrumentation~\cite{Predator}, process-based page protection~\cite{Sheriff}, and the PMU-based sampling~\cite{vtune, Cheetah, 6877463, Laser, Huron, Feather}. Instrumentation-based tools are generally too expensive to be employed in the production environment~\cite{Zhaoqin, Predator}, while Sheriff only supports C/C++ applications using standard synchronizations~\cite{Sheriff}. In theory, Sheriff cannot be able to support some complicated applications (e.g., MySQL) with ad hoc synchronizations. The approaches with the PMU-based sampling is efficient, but with their own shortcomings: Cheetah utilizes a simplified method to compute the number of cache invalidations~\cite{Cheetah}, instead of relying on the sampled cache misses; Jayasena et al. propose a machine learning approach based on the sampled events~\cite{6877463}, Laser utilizes a special type of events (hit-Modified) that may not be available on all hardware~\cite{Laser}, while Feather utilizes the combination of the PMU-sampling and watchpoints to identify false sharing~\cite{Feather}; however, all existing tools typically report an absolute number to evaluate the seriousness of false sharing, which may report insignificant issues. They could not detect false sharing caused by external libraries. 

%\todo{Add tools related to Pin, such as CMP\$im, Cache Pintools, , MultiCacheSim}

%\todo{Check papers citing DProf and cited by DProf}
% Related to cache misses

\paragraph{Classifying Different Types of Cache misses} 
Some approaches could classify multiple types of cache misses together. Sanchez et al. propose a data locality analysis tool that can identify compulsory, conflict and capacity misses, but not coherence misses~\cite{SPLAT}. Its profiling stage incurs reasonable overhead, but it requires a specialized compiler to extract reuse information beforehand and an expensive offline processing stage. These characteristics make this tool inconvenient and inefficient to use. DProf detects datatype-related cache performance issues inside the Linux kernel via the PMU-based sampling and tracing object access histories~\cite{DProf}. DProf employs the definitions of cache misses for its classification, but with the following issues: first, it requires human effort and expertise to summarize data profile, miss classification, working set, and data flow together to identify a particular type of issue, which is not friendly to people without such expertise. Second, it may lose its precision due to its coarse-grained profiling, which is infeasible to find the last write of each miss (and then affect its report). Third, DProf requires the change of the monitored target (e.g., kernel), which may prevent people from using it. Fourth, DProf provides no mechanism of differentiating issues of applications from those of allocators. In contrast, \CP{} overcomes these issues by automatically identifying the type of cache misses, as discussed in Section~\ref{sec:classifier}. Another difference is that \CP{} requires no change of programs, as it is a library that can be linked to applications. Further, \CP{} could also identify cache misses caused by the allocator that DProf cannot do. 

\paragraph{Other Relevant Work:} DMon proposes selective profiling that could incrementally increase its monitoring scope (e.g., sampled events) based on the dynamic behavior of execution~\cite{DMon}. In this sense, \CP{} is very similar to DMon. However, \CP{} selectively chooses the instructions to monitor (not hardware events) in order to collect fine-grained information. DMon relies on \texttt{perf} to collect sampled events, while \CP{} proposes a new profiler that classifies different types of cache misses based on a set of hardware events and hardware breakpoints. \CP{} could classify the type of cache misses, where DMon only reports the cache miss ratio at different lines of code, inheriting from  \texttt{perf}, and relies on human expertise to diagnose the issue. 

%\todo{What type of difference of \CP{}'s sampling events? Precise events? What is the different from the one with MemPerf}

% Not related cache misses
%DaMoN attributes performance issues to different data structures, and proposes the uses of hardware-based sampling to reduce the profiling overhead~\cite{DaMoN}. 
\section{Conclusion}
\label{sec:conclusion}

%As the cache plays a key role in the performance, it is important to design a profiling tool that could report different types of cache miss correctly and efficiently. 
Cache miss is a well-known performance issue. Although existing tools could report cache miss ratios at different lines of code, significant effort is still mandatory to figure out the type and the origin (e.g., object, allocator) of cache misses in order to reduce cache misses.
This paper describes a unified profiling tool--\CP{}-- that could correctly identify different types of cache misses while imposing reasonable overhead.
%differentiate issues of allocators from those of applications, and exclude minor issues without much performance impact. 
This paper further proposes a new method that combines PMU-based coarse-grained sampling and breakpoint-based fine-grained sampling to balance the accuracy and performance.  Overall, CachePerf only imposes 14\% performance overhead, while identifying multiple known and new cache misses correctly. CachePerf is an indispensable complementary to existing profilers due to its uniqueness. 
\begin{acks}
We thank our Shepherd Sergey Blagodurov  and anonymous reviewers for their helpful comments on improving this paper. We also thank Probir Roy, Milind Chabbi for their help on the setup of CCProf and Feather for the comparison, and Xu Liu for the initial discussions on hardware performance counters. This material is based upon work supported by the National Science Foundation under Award CCF-2024253, and the UMass start-up package. Any opinions, findings, and conclusions or recommendations expressed in this material are those of the author(s) and do not necessarily reflect the views of the National Science Foundation.
\end{acks}
{
\bibliographystyle{plain}
\bibliography{ref,tongping, jin, steven}
}

\end{document}